\documentclass[twocolumn]{aastex631}

\renewcommand{\baselinestretch}{1.1}
\usepackage{textcomp}
\usepackage{gensymb}
\usepackage{epstopdf}
\usepackage{amsmath}
\usepackage{graphicx}
\usepackage{hyperref}
\usepackage{multirow}
\usepackage[nameinlink]{cleveref}
\usepackage[multiple]{footmisc}
\usepackage{comment}
\newcommand{\mpl}{M_{\mathrm{p}}}      % planet mass
\newcommand{\mearth}{M_{\oplus}}       % Earth mass symbol
\newcommand{\dd}{\mathrm{d}}
\crefname{paragraph}{\S}{\S\S} % default is {paragraph}{paragraphs}
\defcitealias{kluter:2025a}{Kl{\"u}ter \& Huston et al.}
\shorttitle{RGES Lens Mass \& Distance Measurements} 
\shortauthors{Terry et al.}

\begin{document}

\title{\textbf{Predictions of the Nancy Grace Roman Space Telescope Galactic Exoplanet Survey. IV. Lens Mass and Distance Measurements}}

%\suppressAffiliations

\author[0000-0002-5029-3257]{Sean K. Terry}
\affiliation{Department of Astronomy, University of Maryland, College Park, MD 20742, USA}
\affiliation{Code 667, NASA Goddard Space Flight Center, Greenbelt, MD 20771, USA}

\author[0000-0002-6578-5078]{Etienne Bachelet}
\affiliation{IPAC, Caltech, 1200 E. California Blvd., Pasadena, CA 91125, USA}
\affiliation{Université Marie et Louis Pasteur, CNRS, Institut UTINAM UMR 6213, Besan\c{c}on, France}

\author[0000-0003-2872-9883]{Farzaneh Zohrabi}
\affiliation{Department of Physics and Astronomy, Louisiana State University,
Baton Rouge, LA 70803, USA}

\author[0000-0002-6302-251X]{Himanshu Verma}
\affiliation{Department of Physics and Astronomy, Louisiana State University,
Baton Rouge, LA 70803, USA}

\author[0000-0003-4310-3440]{Alison Crisp}
\affiliation{Department of Astronomy, The Ohio State University, Columbus, OH 43210, USA}

\author[0000-0003-4591-3201]{Macy J. Huston}
\affiliation{Department of Astronomy, University of California Berkeley, Berkeley, CA 94720, USA}

\author[0000-0002-1817-0329]{Carissma McGee}
\affiliation{Department of Aeronautics \& Astronautics, Massachusetts Institute of Technology, Cambridge, MA 02139, USA}

\author[0000-0001-7506-5640]{Matthew Penny}
\affiliation{Department of Physics and Astronomy, Louisiana State University,
Baton Rouge, LA 70803, USA}

\collaboration{20}{(Leading Authors)}

\author[0000-0002-0287-3783]{Natasha S. Abrams}
\affiliation{Department of Astronomy, University of California Berkeley, Berkeley, CA 94720, USA}

\author[0000-0003-3316-4012]{Michael D. Albrow}
\affiliation{University of Canterbury, Department of Physics and Astronomy, Christchurch 8020, New Zealand}

\author[0000-0003-2861-3995]{Jay Anderson}
\affiliation{Space Telescope Science Institute, 3700 San Martin Drive, Baltimore, MD 21218, USA}

\author[0000-0003-4763-066X]{Fatemeh Bagheri}
\affiliation{Code 667, NASA Goddard Space Flight Center, Greenbelt, MD 20771, USA}

\author[0000-0003-0014-3354]{Jean-Philippe Beaulieu}
\affiliation{Sorbonne Universit\'e, CNRS, Institut d’Astrophysique de Paris, IAP, F-75014 Paris, France}
\affiliation{School of Natural Sciences, University of Tasmania, Private Bag 37 Hobart, Tasmania, 7001, Australia}

\author[0000-0003-3858-637X]{Andrea Bellini}
\affiliation{Space Telescope Science Institute, 3700 San Martin Drive, Baltimore, MD 21218, USA}

\author[0000-0001-8043-8413]{David P. Bennett}
\affiliation{Department of Astronomy, University of Maryland, College Park, MD 20742, USA}
\affiliation{Code 667, NASA Goddard Space Flight Center, Greenbelt, MD 20771, USA}

\author[0000-0003-4500-8850]{Galen Bergsten}
\affiliation{Space Telescope Science Institute, 3700 San Martin Drive, Baltimore, MD 21218, USA}

\author[0009-0002-6097-9030]{T. Dex Bhadra}
\affiliation{Department of Astronomy, University of Maryland, College Park, MD 20742, USA}
\affiliation{Code 667, NASA Goddard Space Flight Center, Greenbelt, MD 20771, USA}

\author{Aparna Bhattacharya}
\affiliation{Department of Astronomy, University of Maryland, College Park, MD 20742, USA}
\affiliation{Code 667, NASA Goddard Space Flight Center, Greenbelt, MD 20771, USA}

\author[0000-0002-8131-8891]{Ian A. Bond}
\affiliation{School of Mathematical and Computational Sciences, Massey University, Auckland 0632, New Zealand}

\author[0000-0003-4590-0136]{Valerio Bozza}
\affiliation{Dipartimento di Fisica "E.R. Caianiello", Università di Salerno, Via Giovanni Paolo 132, Fisciano, I-84084, Italy}
\affiliation{Istituto Nazionale di Fisica Nucleare, Sezione di Napoli, Via Cintia, Napoli, I-80126, Italy}

\author{Christopher Brandon}
\affiliation{Department of Astronomy, The Ohio State University, Columbus, OH 43210, USA}

\author[0000-0002-7669-1069]{Sebastiano Calchi Novati}
\affiliation{IPAC, Caltech, 1200 E. California Blvd., Pasadena, CA 91125, USA}

\author[0000-0002-0221-6871]{Sean Carey}
\affiliation{IPAC, Caltech, 1200 E. California Blvd., Pasadena, CA 91125, USA}

\author[0000-0002-8035-4778]{Jessie Christiansen}
\affiliation{IPAC, Caltech, 1200 E. California Blvd., Pasadena, CA 91125, USA}

\author[0000-0003-1827-9399]{William DeRocco}
\affiliation{Department of Astronomy, University of Maryland, College Park, MD 20742, USA}
\affiliation{Department of Physics \& Astronomy, The Johns Hopkins University, 3400 N. Charles Street, Baltimore, MD 21218, USA}

\author[0000-0003-0395-9869]{B. Scott Gaudi}
\affiliation{Department of Astronomy, The Ohio State University, Columbus, OH 43210, USA}

\author[0009-0007-3944-7298]{Jon Hulberg}
\affiliation{Department of Physics, Catholic University of America, Washington, DC 20064, USA}
\affiliation{Code 667, NASA Goddard Space Flight Center, Greenbelt, MD 20771, USA}
\affiliation{Center for Research and Exploration in Space Science and Technology, NASA/GSFC, Greenbelt, MD 20771}

\author[0000-0003-2267-1246]{Stela {Ishitani Silva}}
\affiliation{Code 667, NASA Goddard Space Flight Center, Greenbelt, MD 20771, USA}

\author[0000-0002-7227-2334]{Sinclaire E. Jones}
\affiliation{Department of Astronomy, The Ohio State University, Columbus, OH 43210, USA}

\author[0000-0002-1743-4468]{Eamonn Kerins}
\affiliation{Department of Physics and Astronomy, University of Manchester, Oxford Rd, Manchester M13 9PL, UK}

\author[0000-0002-1910-7065]{Somayeh Khakpash}
\affiliation{Department of Physics, Lehigh University, 16 Memorial Drive East, Bethlehem, PA 18015, USA}

\author[0000-0002-2729-5369]{Katarzyna Kruszy\'nska}
\affiliation{Las Cumbres Observatory, 6740 Cortona Drive, Suite 102, Goleta, CA 93117, USA}

\author[0000-0002-6406-1924]{Casey Lam}
\affiliation{Observatories of the Carnegie Institution for Science, Pasadena, CA 91101, USA}

\author[0000-0001-9611-0009]{Jessica R. Lu}
\affiliation{Department of Astronomy, University of California Berkeley, Berkeley, CA 94720, USA}

\author[0000-0001-5924-8885]{Amber Malpas}
\affiliation{Department of Astronomy, The Ohio State University, Columbus, OH 43210, USA}

\author[0000-0001-9818-1513]{Shota Miyazaki}
\affiliation{Department of Earth and Space Science, Graduate School of Science, Osaka University, Osaka, 560-0043, Japan}

\author[0000-0001-7016-1692]{Przemek Mr{\'o}z}
\affiliation{Astronomical Observatory, University of Warsaw, Al.~Ujazdowskie~4,00-478~Warszawa, Poland}

\author[0009-0004-1245-092X]{Arjun Murlidhar}
\affiliation{Department of Astronomy, The Ohio State University, Columbus, OH 43210, USA}

\author[0000-0001-5825-4431]{David Nataf}
\affiliation{Department of Physics \& Astronomy, University of Iowa, Iowa City, IA 52242, USA}

\author[0009-0002-1973-5229]{Marz Newman}
\affiliation{Department of Physics and Astronomy, Louisiana State University,
Baton Rouge, LA 70803, USA}

\author[0000-0001-8472-2219]{Greg Olmschenk}
\affiliation{Department of Astronomy, University of Maryland, College Park, MD 20742, USA}
\affiliation{Code 667, NASA Goddard Space Flight Center, Greenbelt, MD 20771, USA}

\author[0000-0002-9245-6368]{Radek Poleski}
\affiliation{Astronomical Observatory, University of Warsaw, Al.~Ujazdowskie~4,00-478~Warszawa, Poland}

\author[0000-0003-2388-4534]{Cl\'ement Ranc}
\affiliation{Sorbonne Universit\'e, CNRS, Institut d’Astrophysique de Paris, IAP, F-75014 Paris, France}

\author[0000-0001-5069-319X]{Nicholas J. Rattenbury}
\affiliation{Department of Physics, University of Auckland, Private Bag 92019, Auckland, New Zealand}

\author[0000-0002-9326-9329]{Krzysztof Rybicki}
\affiliation{Astronomical Observatory, University of Warsaw, Al.~Ujazdowskie~4,00-478~Warszawa, Poland}

\author[0009-0000-9738-0641]{Vito Saggese}
\affiliation{Dipartimento di Fisica "Ettore Pancini", Università di Napoli Federico II, Napoli, I-80126, Italy}
\affiliation{Istituto Nazionale di Fisica Nucleare, Sezione di Napoli, Via Cintia, Napoli, I-80126, Italy}

\author[0000-0002-4989-0353]{Jennifer Sobeck}
\affiliation{IPAC, Caltech, 1200 E. California Blvd., Pasadena, CA 91125, USA}

\author[0000-0002-3481-9052]{Keivan G. Stassun}
\affiliation{Department of Physics and Astronomy, Vanderbilt University, Nashville, TN 37235, USA}

\author[0000-0001-8220-0548]{Alexander P. Stephan}
\affiliation{Department of Physics and Astronomy, Vanderbilt University, Nashville, TN 37235, USA}

\author[0000-0001-6279-0552]{Rachel A. Street}
\affiliation{Las Cumbres Observatory, 6740 Cortona Drive, Suite 102, Goleta, CA 93117, USA}

\author[0000-0002-4035-5012]{Takahiro Sumi}
\affiliation{Department of Earth and Space Science, Graduate School of Science, Osaka University, Osaka, 560-0043, Japan}

\author[0000-0002-5843-9433]{Daisuke Suzuki}
\affiliation{Department of Earth and Space Science, Graduate School of Science, Osaka University, Osaka, 560-0043, Japan}

\author[0000-0002-9881-4760]{Aikaterini Vandorou}
\affiliation{Department of Astronomy, University of Maryland, College Park, MD 20742, USA}
\affiliation{Code 667, NASA Goddard Space Flight Center, Greenbelt, MD 20771, USA}

\author[0009-0003-0645-5962]{Meet Vyas}
\affiliation{International Centre for Space and Cosmology, Ahmedabad University, Ahmedabad 380009, Gujarat, India}

\author[0000-0001-9481-7123]{Jennifer C. Yee}
\affiliation{Center for Astrophysics, Harvard \& Smithsonian, Cambridge, MA 02138, USA}

\author[0000-0001-6000-3463]{Weicheng Zang}
\affiliation{Center for Astrophysics, Harvard \& Smithsonian, Cambridge, MA 02138, USA}

\author[0000-0002-9870-5695]{Keming Zhang}
\altaffiliation{NASA Sagan Fellow}
\affiliation{Kavli Institute for Astrophysics and Space Research, Massachusetts Institute of Technology, MA 02139, USA}

\collaboration{51}{(Roman Galactic Exoplanet Survey Project Infrastructure Team)}

\correspondingauthor{S. K. Terry}
\email{skterry@umd.edu}

\begin{abstract}

\small \noindent As part of the Galactic Bulge Time Domain Survey (GBTDS), the Nancy Grace Roman Galactic Exoplanet Survey (RGES) will use microlensing to discover cold outer planets and free-floating planets unbound to stars. NASA has established several science requirements for the GBTDS to ensure RGES success. A key advantage of RGES is \textit{Roman}’s high angular resolution, which will allow detection of flux from many host stars. One requirement specifies that \textit{Roman} must measure the masses and distances of 40\% of detected planet hosts with 20\% precision or better. To test this, we simulated microlensing events toward the GBTDS fields and used Fisher matrix analysis to estimate light curve parameter uncertainties. Combining these with \textit{Roman} imaging observables (lens flux, relative lens-source proper motion), we estimated the achievable precision of lens mass and distance measurements. Using \texttt{pyLIMASS}, a publicly available code for estimating lens properties, we applied this analysis to 3,000 simulated events. Assuming the \cite{cassan:2012a} exoplanet mass function, we find that $\geq$\,40\% of host stars meet the required 20\% precision threshold, confirming that the GBTDS can satisfy the mission requirement. We validated our approach by comparing our inferred lens masses and distances to empirical measurements from detailed image-constrained light curve modeling of historical microlensing events with \textit{Hubble} and Keck follow-up imaging. Our results agree within roughly 1${\sigma}$, demonstrating that both approaches yield consistent and reliable mass and distance estimates, and confirming the robustness of our simulations for Roman-era microlensing science.
\\
\\
\textit{Subject headings}: gravitational lensing: micro, planetary systems, \textit{Roman} \\
\end{abstract}

%-------------------------------------------------------------Introduction-------------------------------------------------------------------------------------------------------------------------------------------

\section{Introduction} \label{sec:intro}
As of 2025, the Nancy Grace Roman Space Telescope (formerly called \textit{WFIRST}; \cite{spergel:2015a}) is undergoing integration and testing at NASA's Goddard Space Flight Center in Greenbelt, Maryland. The telescope is currently scheduled to launch late 2026 or early 2027 and will conduct science operations from a halo orbit around the Sun$-$Earth L2 Lagrange point. There are currently three planned Core Community Surveys (CCS)\footnote{\url{https://www.stsci.edu/roman/surveys-and-programs\#section-d0f7ca7b-88e6-4607-8a76-7218b5082168}} and one planned Early-Definition Astrophysics Survey\footnote{\url{https://asd.gsfc.nasa.gov/roman/comm_forum/forum_21/RGPS_Definition_Committee_Report_01Oct2025.pdf}}. These surveys will exploit the Roman Space Telescope’s unique wide-field, near-infrared, and spectroscopic capabilities to enable transformative investigations across a broad range of astrophysical domains, including the solar system, exoplanetary systems, star formation, Galactic structure, dark matter, and dark energy.\\
\indent The CCS that we focus on in this work is the Galactic Bulge Time Domain Survey (GBTDS). As described in the Roman Observations Time Allocation Committee (ROTAC) Final Report and Recommendations \citep{rotac:2025a}, the GBTDS `overguide' strategy involves high cadence observations during six of the 10 available seasons when the galactic bulge is visible from \textit{Roman}. Each field will be observed every ${\sim}12$ minutes in the wide $F146$ passband, and every few hours in a bluer passband ($F087$) and redder passband ($F213$). During the four seasons in which there are not high cadence observations, \textit{Roman} will visit each of the same fields at a cadence of three days with the $F146$ and $F213$ passbands. Additionally, there will be observations of the survey fields in the five remaining filters that are not used in the high-cadence and low-cadence seasons. These snapshots, which will be important for stellar characterization, will occur at the beginning, middle, and end of each season, for a total of ${\sim}30$ snapshots across the full survey. The Early-Definition Astrophysics Survey we mentioned previously will be the \textit{Roman} Galactic Plane Survey (RGPS), a 700-hour program that will also occasionally observe the GBTDS fields at a significantly lower cadence. Lastly, there will likely be spectroscopic observations of the GBTDS fields with the grism at the beginning, middle, and end of each observing season. These spectroscopic snapshots may allow estimates of stellar temperatures, metallicities, and radial velocities for many of the sources in the GBTDS fields. For a complete description of the GBTDS observing strategy, we refer the reader to the ROTAC report \citep{rotac:2025a}. 

\begin{deluxetable*}{p{0.95\linewidth}}
\tablecaption{\textit{Roman} Level-2 Science Requirements \label{tab:reqs}}
\tablehead{}
\startdata
\textbf{Requirement 1 (EML 2.0.1):} Measure the mass function of exoplanets with $1M_{\oplus} < M < 30 M_J$ and $a\, {\geq}\, 1$ AU to better than 15\% per decade in mass. \\[4pt]
\textbf{Requirement 2 (EML 2.0.2):} Measure the frequency of bound exoplanets with masses in the range $0.1 M_{\oplus} < M < 0.3 M_{\oplus}$ to better than 25\%. \\[4pt]
\textbf{Requirement 3 (EML 2.0.3):} Determine the masses of, and distances to, host stars of 40\% of the detected planets with a precision of $<\,20$\%. \\[4pt]
\textbf{Requirement 4 (EML 2.0.4):} Measure the frequency of free floating planetary-mass objects from Mars to 10 Jupiter masses. If there is one $M_{\oplus}$ free-floating planet per star, measure this frequency to better than 25\%. \\[4pt]
\textbf{Requirement 5 (EML 2.0.5):} Estimate $\eta_{\oplus}$ (frequency of planets orbiting FGK stars with mass ratio and estimated projected semi-major axis within 20\% of the Earth-Sun system) to a precision of 0.2 dex via extrapolation from larger and longer-period planets. \\
\enddata
\end{deluxetable*}

\begin{figure*}[!htb]
\includegraphics[width=1\linewidth]{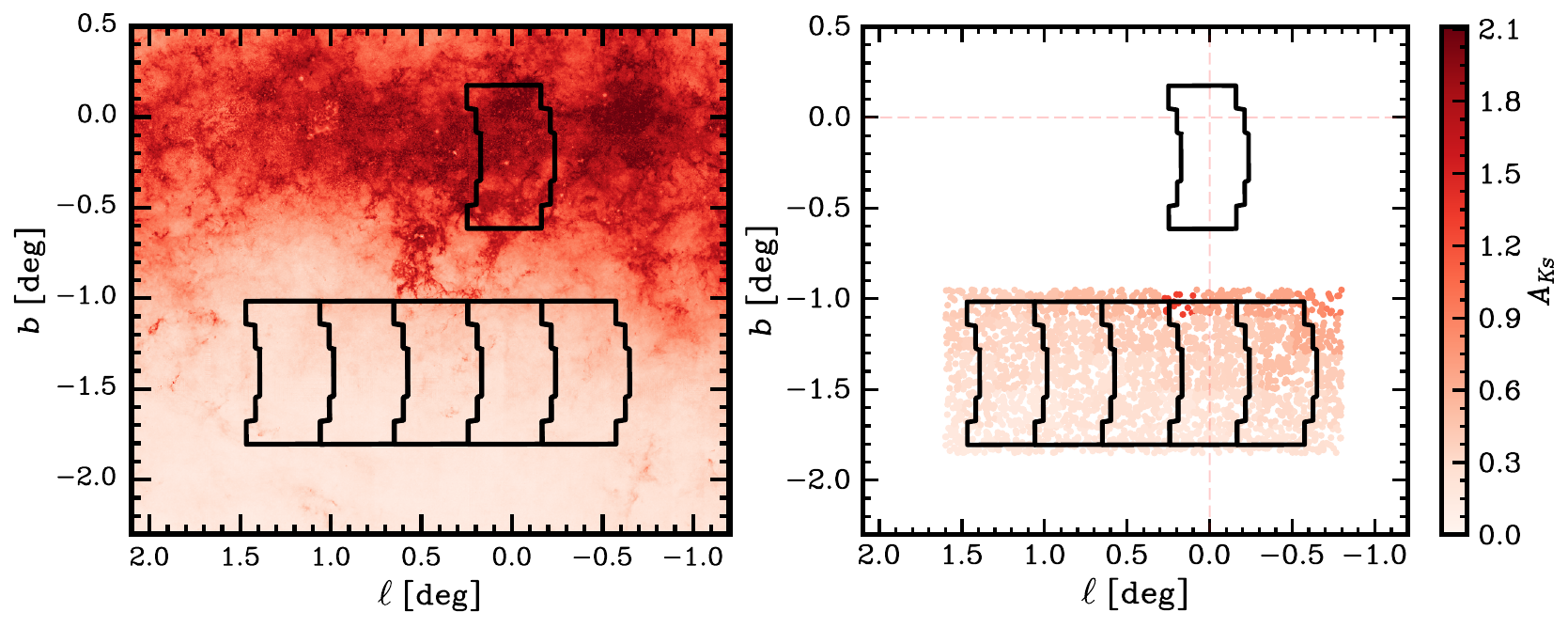}
\centering
\caption{\footnotesize \textit{Left}: GBTDS observational fields (black outlines) overlaid on the $K_S-$band extinction map from \cite{Surot2020}. \textit{Right:} On-sky locations for the selected sample of 3,000 simulated events, colored by $K_S-$band extinction. The color scale between the extinction map and simulated events is identical, therefore the single colorbar can be shared. \label{fig:sim_events_figure}}
\end{figure*}

The left panel of Figure \ref{fig:sim_events_figure} shows the field locations for the GBTDS overguide observing strategy. There are five contiguous fields that are centered at a galactic latitude of $b\, \sim -1.4\degree$ and have relatively low near-IR extinction, and one galactic center (GC) field. We note the field orientations shown in Figure \ref{fig:sim_events_figure} correspond to the northern spring observing seasons, there will be a ${\sim}180\degree$ rotation of the wide-field instrument (WFI) between the Spring and Autumn observing seasons.\\
\indent The most comprehensive study of the bound planet detection rates and expected microlensing yields for the GBTDS comes from \cite{penny19} (``Paper 1" in the current series). Assuming the \cite{cassan:2012a} mass function, \cite{penny19} predict ${\sim}1,400$ bound microlensing planet detections from \textit{Roman} during the full five-year survey. In the years since the \cite{penny19} study, there have been several modifications to the design of \textit{Roman}, the GBTDS observing strategy, our understanding and implementation of Galactic models in population synthesis \citep{kluter:2025a}, and our knowledge of the occurrence rate and mass function of free-floating planets (FFPs) in the Milky Way \citep{gould:2022b, sumi:2023a}. These all have a combined effect on the prior yield estimates. Therefore it is worthwhile to perform an updated analysis of the expected bound planet and FFP yields. \\
\indent The primary goal of the microlensing component of the GBTDS is to `\textit{Carry out a statistical census of planetary systems in the Galaxy, from the outer habitable zone to free floating planets, including analogs to all of the planets in our Solar System with the mass of Mars or greater}'.  In order to ensure the hardware, software, and implemented observations can successfully complete this overall goal, the \textit{Roman} Project defined a set of five high-level science goals.  Three of these goals are aimed at ensuring that, under reasonable assumptions about the frequency of cold exoplanets, the expected yield is sufficient to allow for robust statistical analysis or to place upper limits at the least.  These refer to the yield of cold exoplanets of all masses, the yield of Mars-mass planets specifically, and the yield of FFPs. A fourth goal places a requirement on the ability to constrain the frequency of potentially habitable planets, or $\eta_{\oplus}$.  Forthcoming studies will report the full details of the updated yield estimates of bound and free-floating  planets \citep{zohrabi:inprep}, as well as the expected constraints on $\eta_{\oplus}$ \citep{crisp:inprep}.\\
\indent Here we assess the ability of the GBTDS to meet a fifth requirement, namely that `\textit{the Roman Space Telescope shall be capable of determining the masses of, and distances to, host stars of 40\% of the detected planets with a precision of 20\% or better}'.  Our work relies on the aforementioned paper by \cite{zohrabi:inprep}, as well as an update to the analytic estimates derived by \cite{bennett:2007a} on the uncertainties on the microlensing host star flux and relative proper motion measurements from high-resolution images \citep{verma:inprep}.\\
\indent Our current work is the fourth in a series of papers detailing the expected yields from RGES. \cite{penny19} estimated the yield of several \textit{Roman} architectures for bound planets via microlensing. They used detailed simulations to assess the capabilities of a broad range of Roman mission architectures as well as a range of survey designs, and determined the expected microlensing detection efficiency and planet yield for each configuration. Ultimately, assuming a nominal \textit{Roman} observing strategy of seven total fields observed in six seasons, a wide-passband cadence of ${\sim}15$ minutes, and a fiducial mass function based on the results from \cite{cassan:2012a}, they predict a total yield of ${\sim}1400$ bound exoplanets with mass greater than ${\sim}0.1M_{\oplus}$. \\
\indent The second paper in this series comes from \cite{johnson:2020a}, who estimate the total number of free-floating planet (FFP) detections from \textit{RGES}. They emphasize that the number of FFP detections with \textit{Roman} will depend strongly on the abundance of such FFPs as a function of mass, which is not very well constrained. Adopting the fiducial mass function from \cite{cassan:2012a} that was also used in \cite{penny19}, they predict \textit{Roman} will detect ${\sim}$250 FFPs with masses down to that of Mars.\\
\indent The third study in this series comes from \cite{lastovka:2025a}, who performed a large-scale simulation of planetary microlensing events with exomoons present in the lens systems. They test \textit{Roman}'s ability to detect exomoons of giant planets between $30M_{\oplus}-10M_J$, where this planet mass range was drawn from the \cite{cassan:2012a} mass function. Under the assumption of one moon per microlensing planet, they expect \textit{Roman} to detect of order one exomoon during the nominal five-year GBTDS. We note an older simulation study of microlensing exomoons by \cite{liebig:2010a} shows favorable detectability of these objects in ideal scenarios (e.g. non-giant source stars) with at least 15 minutes cadence and 20 mmag photometric precision, which the GBTDS will easily accomplish. \\
\indent Our current paper is organized as follows: In Section \ref{sec:requirement} we examine the mass and distance requirement and outline our approach to verifying it. In Section \ref{sec:sim_events} we describe the sample of simulated \textit{Roman} microlensing events. In Section \ref{sec:pylimass} we provide an overview of  \texttt{pyLIMASS}, the tool we use to estimate the lens physical properties and uncertainties from the observable parameters of the simulated events. Section \ref{sec:modeling-pylimass} details our comparison of the two primary methodologies that \textit{Roman} will likely employ for microlensing analyses; image-constrained light curve modeling and \texttt{pyLIMASS} Gaussian mixture modeling. Section \ref{sec:conclusion} gives our overall results and includes a discussion on implications for the GBTDS lens mass and distance estimates.

\section{Lens Mass \& Distance Requirement} \label{sec:requirement}

Following the standard NASA procedure, in the mid-2010s, the \textit{WFIRST} microlensing Science Investigation Team worked with NASA through the project to define a set of five science requirements (formally ``Level 2" requirements) to ensure the success of the mission with regards to the microlensing exoplanet demographics survey. These requirements, referred to as EML 2.0.1 through EML 2.0.5 and given in Table \ref{tab:reqs}, define a set of science metrics that are designed to quantify the overall science yield of the GBTDS with regards to the microlensing survey. EML 2.0.1 and 2.0.2 (Requirement 1 and 2 in the Table) define the precision with which the bound planet mass function and frequency of bound Mars mass planets can be measured under the assumption of a modified version of the \cite{cassan:2012a} mass function, as discussed below. These were originally verified in \cite{penny19} and will be re-verified in \cite{zohrabi:inprep} with updated \textit{Roman} hardware parameters and the survey strategy recommended by the ROTAC. EML 2.0.4 (Requirement 4) defines a requirement on the precision with which the frequency of free-floating $1 M_\oplus$ planets can be measured, and this was verified in \cite{johnson:2020a}. EML 2.0.5 (Requirement 5) defines a requirement on the precision with which Roman can measure the frequency of potentially habitable planets ($\eta_\oplus$), and will be verified in \cite{crisp:inprep}. Here we focus on EML 2.0.3 (Requirement 3), which states that the ``Roman Space Telescope shall be capable of determining the masses of, and distances to, host stars of 40\% of the detected planets with a precision of 20\% or better. \\
\indent To place quantitative requirements on the GBTDS, one must make assumptions about the population of exoplanets being probed, including their frequency as a function of mass and semi-major axis. Of course, this frequency is poorly known and is precisely what the survey is designed to constrain. Validating requirements therefore means determining whether \emph{if} a given frequency is assumed the requirement will be met, regardless of the true frequency of planets and thus whether the requirement will in fact be met with the actual survey. At the time when the science requirements were first being formulated, the best estimate of the mass function of cold exoplanets detected by microlensing was that of \cite{cassan:2012a}. \cite{penny19} adopted a modified version of this mass function, assuming that the frequency of planets saturated at two planets per dex below a planet mass of $\mpl < 5.2 M_\oplus$,
{\small
\begin{equation}
\frac{\dd^2 N}{\dd \log \mpl \dd \log a} =
\left\{
\begin{array}{ll}
0.24\ \text{dex}^{-2} \left(\frac{\mpl}{95\mearth}\right)^{-0.73} & \text{if } \mpl \ge 5.2\mearth,\\
2\ \text{dex}^{-2} & \text{if } \mpl < 5.2\mearth,
\end{array}
\right.
\label{fiducialmf}
\end{equation}
}

\noindent where $a$ is semi-major axis, $\mpl$ is planet mass, and $N$ is the average number of planets per star. 
Note that this is both a mass function and semi-major axis distribution, with the distribution of planets semi-major axis assumed to be uniform in $\log a$.\\
\indent This form was used to define and verify science requirements EML 2.0.1, 2.0.2 \citep{penny19} and 2.0.4 \citep{johnson:2020a}, and will be used to verify EML 2.0.5. We note it was also used for the population of host planets for evaluating the yield of exomoons orbiting wide-separation planets from Roman by \cite{lastovka:2025a}. While there have since been many microlensing analyses which derive an updated MF or mass ratio function (e.g. \cite{suzuki:2016a, suzuki:2018a}, \cite{bennett:2021a}, \cite{zang:2025a}), it is essential that, when verifying the science requirements, we adopt the same assumptions as were used to define these requirements.  We therefore adopt this modified version of the \cite{cassan:2012a} MF when verifying EML 2.0.3 (Requirement 3) on the host star masses and distances (see Section \ref{subsec:pylimass_results}).

\subsection{Light Curve Higher-Order Effects} \label{subsec:high-order-effects}

Following the formalism of \cite{gould:2000b}, we can define the angular Einstein radius ($\theta_E$) as the angular radius of the Einstein ring as seen by an observer. It can be expressed as follows:

\begin{equation}
    \theta_E \equiv \sqrt{\kappa M_L \pi_{\rm rel}}
\end{equation}

\noindent where $M_L$ is the mass of the lens, $\kappa = 4G/c^2$AU, and $\pi_{\rm rel} \equiv \pi_L - \pi_S$ is the relative lens-source parallax. It is also useful to define the microlensing parallax vector $\vec{\pi}_E$,  which has a magnitude 
\begin{equation}
    \pi_E = \frac{\pi_{\rm rel}}{\theta_E},
\end{equation}
and direction equal to that of the relative lens-source proper motion $\vec{\mu}_{\rm rel}$. Lastly, the relative lens-source proper motion has a magnitude equal to $\mu_{\rm rel} = \theta_E/t_E$ (see also Eq. \ref{eq:murel_imaging}).\\
\indent While the above formulae are defined for a single lens and single source, typical binary light curve modeling includes additional parameters $q$, the mass ratio between the two lens components and $s$, the separation between the two lens components, in units of the angular Einstein radius, $\theta_E$. The mass ratio and separation are almost always measured with reasonable precision, although there are well-known degeneracies that can make determining the true parameters difficult (e.g., \cite{dominik:1999a}, \cite{yee:2021a}, \cite{zhang:2022a}). The Einstein radius crossing time, $t_E$, is the time it takes for the lens-source relative motion to traverse the Einstein radius ($\theta_E$). This parameter is also typically well-measured, however in cases of faint sources with very high magnification planetary anomalies, it can be difficult to measure $t_E$ precisely because this parameter must be measured from the low-magnification region of light curves \citep{alard:1997a}. Another typical light curve modeling parameter is the de-reddened source star magnitude and color (e.g. $I_{S,0}, (V_{S,0}-I_{S,0})$). This can be used to estimate the source star angular radius, $\theta_*$, using the dereddened flux of the source and color-surface brightness relations \citep{kervella:2004a, boyajian:2014a, adams:2018a}. \\
\indent While some of these routine observables, namely the dimensionless quantities $q$ and $s$, provide insight into the properties of the detected planetary systems, it is generally quite valuable if physical quantities can also be measured, such as the planet mass $\mpl$ (in, e.g,. $M_\oplus$ or $M_J$), the host star mass $M_L$ (in $M_\odot$), the semi-major axis of the planet $a$ (in au), as well as the distance to the system $D_L$ (in pc or kpc). Importantly, when determined as a function of these parameters, microlensing demographic constraints can then be directly compared to those determined with other methods. Generally, translating the routine light curve observables to physical parameters requires a measurement of the host mass $M_L$ and distance $D_L$.  We note that, with a measurement of these two parameters, it is possible to express $s$ in physical units, i.e., the instantaneous projected separation $a_\perp= s \theta_{\rm E} D_L$.  However, the true semi-major axis is related to $a_\perp$ through the phase, inclination, eccentricity, and argument of periastron of the orbit, which are generally unknown.\\
\indent In order to achieve direct mass measurements of the host and companion using only information from the microlensing light curve itself, one must precisely measure two quantities from higher-order effects in the microlensing light curve. Finite source effects occur when the light curve deviates from a point-source magnification due to the finite size of the source in regions where the magnification changes quickly over the size of the source. This effect is prominent for sources that approach within a few source radii from a caustic curve. The magnitude of finite source effects depends on the angular size of the source ($\theta_*$), in units of
the angular Einstein ring radius ($\rho$), where $\rho = \theta_* / \theta_E$. Further, the source radius crossing time, $t_*  \equiv \rho\, t_E$, is another measure of the finite source effect. Many planetary microlensing events to date have had $t_*$ measured reasonably well, which allows another way to measure the angular Einstein radius; $\theta_E = t_E\,\theta_*/t_*$. Notice a measurement of the finite source effect allows one to infer $\theta_E$, since $\theta_*$ can typically be inferred from the source flux and color with a measurement or assumption of the extinction. As mentioned previously the microlensing parallax, $\pi_{\rm E}$, can be measured by a distortion in the light curve due to the acceleration of the observer, typically due to the orbital motion of the Earth \citep{alcock:1995a}. Microlensing parallax is always present in a microlensing event, but not always measurable. If measured, usually the East component is more tightly constrained than the North component because the orbital acceleration of the Earth perpendicular to the line-of-sight to the galactic bulge is largely in the East-West direction. \\
\indent A measurement of both finite source effects and microlensing parallax is very powerful, because both provide essentially a geometric constraint on the mass and distance to the lens by comparing the angular Einstein ring radius projected to the source plane or observer plane to a `ruler’ in that plane,  where the rule is either angular size of the source or the size of the Earth’s orbit. Thus, the measurement of each quantity can be used to create a relation between $M_L$ and $D_L$. Importantly, because they are projected to opposite planes, these relations are `inverted’, such that, if both $\theta_{\rm E}$ and $\pi_{\rm E}$ are measured, the relations can be solved to give an unambiguous measurement of $M_L$ and $D_L$. The equations that relate these two quantities to the lens mass and distance are given in \cite{bennett:2008a} and \cite{gaudi:2012a}:
\begin{equation}\label{eq:ml_rho}
    M_L = \frac{\theta_E^2}{\kappa \pi_{\rm rel}}= \frac{c^2}{4G}\theta_E^2\frac{D_S D_L}{D_S - D_L},
\end{equation}
and
\begin{equation}\label{eq:ml_pie}
    M_L = \frac{\pi_{\rm rel}}{\kappa \pi_E^2}=\frac{c^2}{4G}\frac{\textrm{AU}^2}{{\pi_E}^2}\frac{D_S - D_L}{D_S D_L}
\end{equation}
\noindent Therefore, equations \ref{eq:ml_rho} and \ref{eq:ml_pie} can be combined to yield the lens mass, with no dependence on the lens or source distances:
\begin{equation}\label{eq:ml}
    M_L = \frac{\pi_E}{\kappa \theta_E}=\frac{c^2\textrm{AU}}{4G}\frac{\theta_E}{\pi_E}.
\end{equation}
Similarly, the relative lens-source parallax is given by
\begin{equation}\label{eq:pirel}
    \pi_{\rm rel} = \theta_E\pi_E,
\end{equation} 
and the lens distance is given by
\begin{equation}\label{eq:dl-ls}
    D_L^{-1} = \frac{\theta_E\pi_E}{\textrm{AU}} + D_S^{-1}.
\end{equation}

\indent Thus, when both $\theta_E$ and $\pi_E$ are well-measured, it is possible to measure the mass of the lens, and with an estimate or measurement of the distance to the source, calculate the lens distance as well and thus obtain complete solution to a properties of the microlensing lens system in physical units. This is largely driven by the fact that unambiguous measurements of $\pi_E$ and $\theta_E$ lead to a geometric measurement of $M_L$ and $D_L$, which avoids dependencies on somewhat complex astrophysics (metallicity-dependent, extinction, stellar atmosphere, etc). One exception is the inference of $\theta_*$ from the flux and color of the source, which depends on knowing the extinction and the color-surface brightness relation. This relation requires accurate and empirical-based calibrations using known stars, typically of solar-type \citep{boyajian:2014a}. \\
\indent Historically, it has been rare for both of these higher-order effects to be precisely measured in microlensing light curves. Less than two dozen published microlensing events have had complete solutions for the lens systems measured in this way (see, e.g. \citealt{han:2025a} and references). The vast majority of published planetary microlensing events rely on Bayesian analyses of the events that assume host stars of all possible masses have an equal probability to host a planet with the measured mass ratio, $q$. This prior assumption is not based on any empirical data, in fact there is preliminary evidence from microlensing and radial velocity surveys that suggest the planet hosting probability scales in proportion to the host star mass \citep{nunota:2024a, bennett:inprep}. It is  advisable to avoid priors that might be overly prescriptive, as these may interfere with unexpected discoveries in the era of very-high precision photometry and astrometry from \textit{Roman}. A full accounting and description of the light curve higher-order effects included and measured in our simulated \textit{Roman} events (Section \ref{sec:sim_events}) can be found in \cite{zohrabi:inprep}.

\subsection{Measuring Lens Flux with \textit{Roman}} \label{subsec:lens-flux-description}
In addition to the higher-order effects that can be measured from the light curve model fitting, there is a third quantity that can yield a mass-distance relation. A measurement of the lens star flux in a given passband ${\cal F}_\lambda$ can be used alongside a mass-luminosity relationship in that passband (either from empirical mass-luminosity relations \citep{henry:1993a, kenyon:1995a, henry:1999a, delfosse:2000a} or theoretical isochrones \citep{girardi:2002a, bressan:2012a}) to yield a third lens mass-distance relation
\begin{equation}\label{eq:md-3}
    {\cal F}_\lambda = \frac{L_\lambda(M_L)}{4\pi D_L^2} \ 
\end{equation}
\noindent Here $L_\lambda(M_L)$ is the relationship between the mass of a star and its specific luminosity, i.e., its luminosity in a given filter or passband. It is worthwhile noting that this relationship depends not only on the specific passband, but in principle also depends on other properties of the star such as its metallicity, surface gravity, age, etc., which are generally unknown.  Isolating the flux from the lens typically requires high-resolution imaging of the target(s) to disambiguate any nearby unrelated and/or blended stars from the true microlensing target. An important fitting parameter in the microlensing modeling is the source flux ($f_S$) in units of flux or magnitude in a given filter. One can then compare the measurement of the target brightness in the high-resolution imaging to that of the source brightness from the light curve model. The difference in these two brightnesses is often referred to as ``excess flux" at the position of the source, which can be attributed to the lens star, a companion to the source or lens, or an unrelated ambient field star that happens to be blended with the target(s). Excess flux measurements have been made for several microlensing targets in the past; \cite{bennett:2010b, kubas:2012a, koshimoto:2017b}. \\
\indent As we just mentioned, there is some ambiguity in the interpretation of excess flux at the position of the source star. Although high-resolution follow up imaging typically occurs shortly after a microlensing event when the source magnification is back to its baseline brightness, the lens and source are still very highly blended and cannot be independently measured. At this epoch it is difficult to definitively conclude that all of the measured excess flux comes entirely from the lens star. On the other hand, there is another strategy referred to as ``late-time" high-resolution follow up imaging which typically occurs multiple years after a microlensing event. This allows time for the source and lens to separate on the sky, enabling them to be independently measured. This largely mitigates the issue of ambiguous excess flux attribution, particularly if the lens-source measurement is consistent with the prediction from the relative proper motion inferred from the light curve alone from $t_*$ when finite source effects are robustly detected.\\
\indent There are two additional relations between the lens-source relative proper motion and the angular Einstein ring radius that are of great importance:

\begin{equation}\label{eq:murel_lightcurve}
    \mu_{\textrm{rel,G}} = \frac{\theta_*}{t_*},
\end{equation}

\noindent and

\begin{equation}\label{eq:murel_imaging}
    t_E = \frac{\theta_E}{\mu_{\textrm{rel,H}}},
\end{equation}

\noindent where $t_E$ is the Einstein ring crossing time (in days), the typical timescale of a microlensing event. Equation \ref{eq:murel_lightcurve} represents the relative proper motion estimate that comes from the light curve. It is typically evaluated in a geocentric reference frame, hence the subscript $G$ \citep{gould:2004a}. In contrast, the relative proper motion from a direct measurement of the lens-source separation by \textit{Keck} or \textit{HST} (or \textit{Roman} in the future) is in the Heliocentric reference frame, hence the $H$. We briefly note that \textit{Roman} will naturally conduct its own `follow-up' imaging of events that it detects, therefore it is feasible that the relative proper motion from \textit{Roman} light curve-only data and \textit{Roman} follow-up imaging data will be in the same reference frame (\textit{Roman-centric}). The relation between Geocentric and Heliocentric relative proper motion is given in \cite{dong:2009b}:

\begin{equation}\label{eq:mu-rel}
\vec{\mu}_{\textrm{rel,H}} = \vec{\mu}_{\textrm{rel,G}} + \frac{{\vec{v}_{\Earth}}{\pi_{\textrm{rel}}}}{AU} \ ,
\end{equation}

\noindent where $\vec{v}_{\Earth}$ is Earth's projected velocity relative to the Sun at the time of peak magnification in the Geocentric reference frame and $\pi_{\rm{rel}} \equiv \rm{AU}/D_{L} - \rm{AU}/D_{S}$. For distant lenses ($> \,4$ kpc) toward the galactic bulge, the difference in relative proper motion between these reference frames is negligible.\\
\indent For observations taken at time $\Delta t$ after a microlensing event, the proper motion uncertainty for unblended stars degrades as $(\Delta t)^{-1}$. The uncertainties in either the flux ratio or the separation $\Delta x$ between two partially blended stars when the other is known depends on $(\Delta x)^{-2} \propto (\Delta t)^{-2}$, and depends on $(\Delta x)^{-3} \propto (\Delta t)^{-3}$ when neither are known \citep{bennett:2007a}. This means for observations of microlensing systems detected by \textit{Roman}, a mixture of immediate imaging (e.g., while the microlensing event is ongoing), imaging soon after the exoplanet is detected (e.g., when any stars blended with the host star have not moved substantially), and imaging well before or well after the exoplanet is detected (when the host and blended star are sufficiently separated) will be most useful. \textit{Roman} itself is poised to make one or more of these time-sensitive observations. The GBTDS is nominally scheduled to span a 5-year baseline, which will allow for many of the lens-source pairs to separate sufficiently such that one may extract direct lens flux measurements. This will enable \textit{Roman} to conduct its own ``early-time" or ``late-time" high-resolution imaging for microlensing events that are detected in any of the six-season GBTDS.  \\
\indent There are two primary regimes to consider when the lens and source are separated by some amount, $\Delta x$, but still highly blended (e.g. not separately resolved): The first regime occurs when the dominant observable is the image elongation. This happens when the separation between the source and lens, $\Delta x$, is smaller than but comparable to the full width at half maximum (FWHM) of the point spread function (PSF). The second regime arises when the primary observable is the color-dependent centroid shift (CDCS). In this case, $\Delta x$ is sufficiently smaller than the PSF’s FWHM such that image elongation within any single passband is negligible, yet large enough that the difference in the centroids of the PSFs in two passbands—caused by the differing colors of the source and lens—is detectable. Many prior studies have measured both of these effects with Keck adaptive optics (AO) and \textit{HST} high-resolution imaging (\cite{bennett:2006a, batista:2015a, bennett:2015a, bhattacharya:2018a, terry:2021a, bhattacharya:2021a, terry:2022a, bennett:2024a, terry:2024a, vandorou:2025a, vandorou:2025b}).

\subsection{Measuring Lens-Source Proper Motion with \textit{Roman}} \label{subsec:murel_description}
\noindent The work of \cite*{bennett:2007a} derived several analytic relations that describe the ability to detect lens (host) stars and measure relative proper motions ($\mu_{\textrm{rel}}$) in blended lens-source pairs with high-resolution \textit{HST} imaging. In general, luminous lens stars would typically not be fully resolvable in high-resolution (\textit{HST}) images taken less than a decade after peak magnification. However, with a reasonably stable PSF, it is possible to measure lens-source separations that are much smaller than the width of the PSF FWHM. From its location at L2, we expect \textit{Roman's} PSF stability to be exquisite, enabling many lens flux detections and $\mu_{\textrm{rel}}$ estimates through CDCS and/or image elongation.\\
\indent The ability to measure the elongation/CDCS in a given PSF depends on the total photon counts ($N_{\rm tot}$), angular separation between the lens and source ($\beta$), and the the fraction of the total (lens+source) flux contributed by the lens ($f_l$). We assume the primary observations will be taken in \textit{Roman}'s $F146$ filter with an exposure time of 66.9 seconds and a cadence of 12.1 minutes, with secondary observations replacing these primary observations every 6 hours in each of the $F087$ and $F146$ filters. In practice, the image elongations and centroid shifts will be measured from analysis of stacked, oversampled images. These will be created using ${\sim}$1 week of single epoch images in the primary F146 filter, corresponding to ${\sim}750$ individual images and a total exposure time of 14.55 hours, and 72 days of images in the secondary filters, corresponding to 576 images and a total exposure time of 10.70 hours. The total photons ($N_{\rm tot}$) in the PSF of the blended source and lens is then determined using the source and lens magnitude for each simulated event.\\
\indent Under the assumption of a Gaussian PSF model with standard deviation specified by the Roman detector\footnote{\url{https://roman.gsfc.nasa.gov/science/WFI_technical.html}}, we estimate the fractional error in the relative proper motion $\mu_{\rm rel}$ via the following analytic relation,
\begin{equation}\label{eq:image_elong}
    \frac{\sigma_{\mu_{\rm rel}}}{\mu_{\rm rel}} \approx 0.02 \sqrt{\frac{10^7}{ N_{\rm tot}}} \left[\frac{\sigma_\lambda}{45~{\rm mas}} \frac{5 ~{\rm mas}}{\beta}\right]^2 \frac{1}{f_l(1-f_l)},
\end{equation}
where we take the standard deviation $\sigma_\lambda$ based on the chosen filter. The error in the proper motion is calculated using the baseline $\beta = \max(T_{\text{max}} - t_0, t_0) \cdot \mu_{\text{rel}}$, Where $t_0$ is time of the peak of the event from the start of the first GBTDS season, $T_{\rm max} = 2010)$ days is the maximum duration of the GBTDS campaign. This then assumes the image elongation and centroid shift is measured from the two individual stacks that have the largest separation in time. However, we note that in practice there is additional information in the intermediate stacks that we have not considered, and thus our estimates are conservative.\\
\indent We also estimate the fractional error on the relative proper motion, $\sigma_{\mu_{\rm rel}}/\mu_{\rm rel}$, using the CDCS method, which measures the displacement of the centroid between different filters, arising from the distinct spectral energy distribution of the source and lens (e.g. different color stars). For each filter combination, we compute the error in proper motion by propagating the error in measuring the centroid in the two filters which is given by:
\begin{align}\label{eq:cdcs}
    \frac{\sigma_{\mu_{\rm rel}}}{\mu_{\rm rel}} &\approx 0.003
\left[\left(\frac{\sigma_{\lambda_1}}{45~{\rm mas}}\right)^2\frac{10^7}{N_{\rm tot,1}} + \left(\frac{\sigma_{\lambda_2}}{31~{\rm mas}}\right)^2\frac{10^7}{N_{\rm tot,2}} \right]^{1/2} \nonumber\\
&\frac{5~{\rm mas}}{\beta} \frac{1}{|f_{l_1}-f_{l_2}|},   
\end{align}

\noindent where $f_{l_1}$ and $f_{l_2}$ are the lens flux fraction in the two passbands with PSF size $\sigma_{\lambda_1}$ and $\sigma_{\lambda_2}$ respectively. In this case, we use the baseline $\beta= T_{\text{max}} \cdot \mu_{\text{rel}}$ representing the full observational duration since the centroid shift signal accumulates continuously throughout the GBTDS observation period.\\
\indent In this work we assume a negligible background. This assumption can be somewhat problematic because in real \textit{Roman} data we expect there will be a measurable contribution to the noise in the PSF due to the background. We refer the reader to \cite{verma:inprep} for further simulation work that incorporates background noise into \textit{Roman} PSFs. The combination of elongation and CDCS measurements across multiple filters provides robust constraints on the microlensing geometry, and for each simulated event we select the filter combination and measurement method which give the minimum error on $\mu_{\textrm{rel}}$. For various remnant lenses (white dwarfs,  neutron stars, and black holes), as well as brown dwarfs, there is no appreciable flux to measure, therefore both CDCS and image elongation methods will always fail to provide any estimate of lens flux and $\mu_{\textrm{rel}}$. Our sample of simulated events include some groups of `dark' lenses, which are described in Section \ref{sec:bdwds}. Finally, we have only briefly discussed the CDCS and image elongation methodology here, a forthcoming paper will present a full analysis and a major update to the \cite*{bennett:2007a} analytic relations for \textit{Roman} \citep{verma:inprep}.

\subsection{Euclid and HST Precursor Imaging of the GBTDS Fields}
There have recently been two dedicated campaigns\footnote{\url{https://www.cosmos.esa.int/web/euclid/egbs}}\textsuperscript{,}\footnote{\url{https://archive.stsci.edu/proposal_search.php?mission=hst&id=17776}} to obtain high-resolution imaging of the GBTDS fields several years before the start of \textit{RGES}. These pre-imaging programs have a shared goal of increasing the time baseline of high-resolution imaging for events in the lower-extinction GBTDS fields, which enables a higher fraction of lens detections and direct lens-source relative proper motion measurements in future \textit{Roman}-detected microlensing events.\\
\indent The prospect of \textit{Euclid} to deliver microlensing demographics has been considered in several past studies \citep{beaulieu:2010a, penny:2013a}. With the launch of \textit{Euclid}, additional work has been conducted that includes studying how \textit{Roman} microlensing mass measurements can be improved with early \textit{Euclid} observations \citep{bachelet:2022a}. Further, \cite{bachelet:2019a} and \cite{beaulieu:2021a} investigated science cases that can be optimized with simultaneous \textit{Euclid} and \textit{Roman} observations later this decade. In March 2025, \textit{Euclid} obtained high-resolution images covering 5.2 deg$^2$ of the inner bulge south of the GC (including the GBTDS footprint) with the \textit{VIS} Camera. Compared to the regular \textit{Euclid} observing sequence, the exposure time was slightly shorter (400 sec instead of 550 sec), with 16 dithers at each of the nine total pointings to ensure a well-sampled PSF. This program is known as the \textit{Euclid} Galactic Bulge Survey (s). Additional calibration fields were observed under the same thermal environment before and after the EGBS, which ensures a proper PSF modeling. The data is currently undergoing processing by the \textit{Euclid} Exoplanet Science Working Group, with a scheduled public release in June 2026. \\
\indent The large \textit{HST} program GO-17776 \citep{terry:2025b} is currently obtaining high-resolution images with both the Wide-Field Camera 3 (\textit{WFC3}) and Advanced Camera for Surveys (\textit{ACS}) in coordinated-parallel imaging mode. As mentioned previously, much like the EGBS, one primary goal of this \textit{HST} precursor survey is to extend the time baseline for which \textit{Roman} lenses and sources can be studied with high-resolution imaging. The \textit{HST} survey is taking parallel images in identical filter sets for both imagers, the $F814W$ filter (wide $I-$band) and the $F606W$ filter (wide $V-$band). While the EGBS took approximately 24 hrs to complete its survey which covered the full lower-extinction fields of the GBTDS, the significantly smaller \textit{HST} FoV has required more on-sky time, and covers only ${\sim}$65\% of the lower-extinction GBTDS footprint. The \textit{HST} survey began in early 2025, completing 70\% of the total imaging during the 2025 galactic bulge observing season. The overview, strategy, and first results from this dataset will be presented in \cite{terry:inprep}. The remaining 30\% of the survey is scheduled to be conducted in early 2026, when the bulge becomes observable from \textit{HST}. \\
\indent The inclusion of these two precursor datasets will synergize well for the microlensing science cases during \textit{Roman}, particularly for lens mass and distance measurements. A joint analysis of the the well-dithered \textit{Euclid} VIS images with the multi-passband \textit{HST} images will effectively raise the number of \textit{Roman}-detected microlensing events that have successful lens detections and relative proper motion measurements. Lastly, although this will be a significant benefit for the GBTDS as a whole, we do not consider the inclusion of these precursor datasets in our current analysis because the science requirements (Table \ref{tab:reqs}) do not rely on precursor observations.

\begin{deluxetable}{lc}[!htp]
\deluxetablecaption{Simulated Survey Parameters \label{tab:survey-params}}
\tablecolumns{2}
\setlength{\tabcolsep}{5pt}
\tablewidth{\columnwidth}
\tablehead{
\colhead{\hspace{-2.5cm}Parameter} & \colhead{\textit{Roman GBTDS}}
}
\startdata
Total area [deg$^2$] & 1.8\\
Total Seasons & 6\\
Total Fields & 6\\
Season Length [day] & $72$ \\
Primary filters & $F087, F146, F213$ \\
Exposure time [sec] & 66.9\\
$F146$ Cadence [min] & 12.1\\
$F087,\, F213$ Cadence [hr] & 6.0\\
Total $F146$ exposures & ${\sim}$8,400 season$^{-1}$\\
Total $F087,\, F213$ exposures & ${\sim}$300 season$^{-1}$\\
Photometric precision & 0.02 mag $@\, F146\,{\sim}\,21.2$\\
\enddata
%\tablenotetext{}{\footnotesize{$^{\dagger}$ $F087$ and $F213$ are likely to be used filters, but are not yet officially selected by the \textit{Roman} project.}}
\end{deluxetable}

%------------------Simulated Events----------------------------------------------------------------------------------------------------------------------------
\section{Simulated \textit{Roman} Microlensing Events} \label{sec:sim_events}

As mentioned in Section \ref{sec:intro}, the previous simulations of \cite{penny19} were conducted using a previous version of the \texttt{GULLS} software suite \citep{penny:2013a}. \texttt{GULLS} simulates individual microlensing events with source and lens stars that are drawn from an underlying galactic model. Each simulated event is assigned a normalized weight ($w_i$) that is proportional to its contribution to the total microlensing event rate for a given sight line. As \cite{penny19} point out, their simulations are subject to various uncertainties including; certain specifications of the telescope design and detector response during in-flight science operations, the ability to accurately measure and model the astrophysical components that govern the microlensing events (e.g. the Galactic model and stellar populations), and uncertainty in the underlying population of bound- and free-floating planets and how those affect the yields that they ultimately predict. In particular, the version of \texttt{GULLS} used in \cite{penny19} drew lenses and source from a version of the Besan\c{c}on galactic model \citep{robin:2003a} with an ad hoc correction to the event rates. Furthermore, \cite{penny19} adopted a nominal survey design that differs significantly from the current survey design recommended by the ROTAC \cite{rotac:2025a}, which also impacts the yields.\\
\begin{figure*}[!htb]
\includegraphics[width=0.9\linewidth]{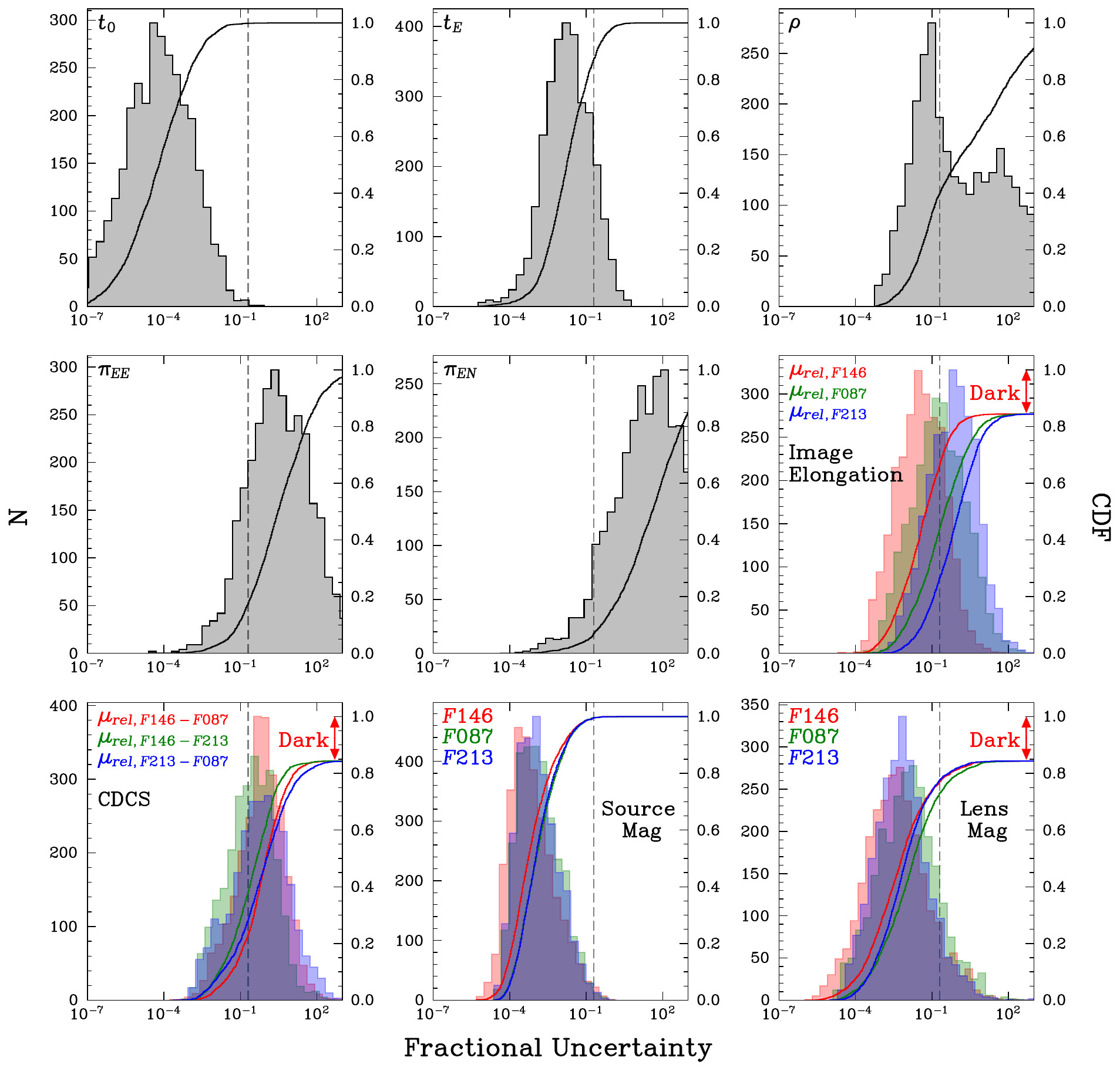}
\centering
\caption{The distributions of fractional uncertainty on observables for 3,000 simulated events processed in \texttt{pyLIMASS}. Fisher errors were derived for parameters $t_0, t_E,\rho,\pi_{EE},$\, and $\pi_{EN}$ (grey histograms). Errors on $\mu_{\textrm{rel}}$ were derived via Eq. \ref{eq:image_elong} and \ref{eq:cdcs}. Errors on source and lens magnitudes were derived from the baseline and source flux fraction constraints via light curve fits (see \cite{zohrabi:inprep}). The cumulative distribution functions (CDFs) for each parameter are overlaid as solid curves, and we show a dashed vertical line which corresponds to a fractional uncertainty of 0.2. For the image elongation, CDCS, and lens magnitude distributions there is a gap in the CDF between ${\sim}0.85$ and 1.0, which is due to the contribution of dark lenses that emit effectively zero flux in these passbands and thus do not provide any constraint on the proper motion via image elongation or CDCS. We note the extinction is also given as an input observable in \texttt{pyLIMASS}, however it has a fixed error of $\sigma A_K = 0.1$ for all events. \label{fig:uncertainty_grid}}
\end{figure*}
\indent In order to partially address these uncertainties and updates, \cite{zohrabi:inprep} has performed an updated simulation of microlensing events based on new specifications of the \textit{Roman} detectors and overall GBTDS observing strategy. \cite{zohrabi:inprep} have also implemented the \texttt{SynthPop} Galactic model \citep{kluter:2025a} within \texttt{GULLS}, which is a modular population synthesis code that pulls from existing models and observational work (c.f. \citealt{koshimoto:2021a}). The simulations closely follow the GBTDS Core Community Definition Committee recommended `Overguide' observing strategy. Table \ref{tab:survey-params} reports the survey parameters that were used to generate the simulated microlensing events that we subsequently analyze in this work. We note that although microlensing events were simulated toward the GC via \texttt{GULLS}, we do not consider any simulated events in the field toward the GC (see right panel of Figure \ref{fig:sim_events_figure}), as the original level-2 requirement on lens mass and distance estimates was referenced to the lower-extinction fields that are the primary focus of the microlensing survey.  Although they will be scientifically interesting, we do not include events in the GC field as they would impact final statistics and thus influence our conclusions as to \textit{Roman}'s ability to meet this requirement.\\
\indent Over 500,000 planetary microlensing events were simulated using \texttt{GULLS}. The simulated events were split into six distinct planetary mass bins; $0.1, 1, 10, 100, 1000, 10000M_{\oplus}$ planets. These mass bins correspond to planetary mass ranges that are explicitly referenced in several level-2 science requirements\footnote{\url{https://rges-pit.org/about/}}. The planet semi-major axes were drawn from a uniform distribution in log space between 0.3 and 30 AU in order to represent the full range of possible orbital separations. The caustic region of influence (CROIN) parameterization \citep{penny:2014a} was adopted for computational efficiency. The full details of the simulations will be presented in a forthcoming work by \cite{zohrabi:inprep}.

\begin{deluxetable}{lc}[!htb]
\deluxetablecaption{Simulated Catalog Cuts \label{tab:cutting-thresholds}}
\tablecolumns{2}
\setlength{\tabcolsep}{6pt}
\tablewidth{\columnwidth}
\tablehead{
\colhead{\hspace{-1.4cm}Cut Performed} & \colhead{Events Remaining}
}
\startdata
All Events & 511,901\\
$(\chi^2_{\textrm{1L1S}} - \chi^2_{\textrm{2L1S}}) > 160$ & 233,731\\
$t_0$ in season & 173,739\\
$|u_0| < 3$ & 134,504\\
5-field GBTDS Footprint & 61,511\\
\hline
\hline
Final Selection & \\
\hline
$p_{i,b} = \dfrac{w_i}{\sum_{b=1}^{N_b} w_i}$
 & 3,000\\
\enddata
%\tablenotetext{}{}
\end{deluxetable}

%-------------------Sample Selection-----------------------
\subsection{Sample Selection} \label{sec:sample_select}
One of our main goals in this work is to process a statistically robust set of simulated events through the \texttt{pyLIMASS} software \citep{bachelet:2024a}. We describe \texttt{pyLIMASS} in more detail in Section \ref{sec:pylimass}. In order to prepare a realistic set of events that is also tractable, we first make some relatively straightforward cuts to the initial sample of 500,000$+$ objects. Table \ref{tab:cutting-thresholds} gives an overview of our cutting criteria and final selection that we make. First, we select events for which the $\chi^2$ for the binary-lens single-source models are favored over the single-lens single-source models by at least $\Delta \chi^2 > 160$. The choice of 160 is relatively arbitrary. However, similar $\Delta \chi^2$ thresholds have been used in statistical studies of planetary microlensing events in the past \citep{suzuki:2016a, suzuki:2018a, koshimoto:2023a, zang:2025a}. Our next cutting threshold is on the time of peak magnification, $t_0$. We restrict $t_0$ to occur within one of the six GBTDS seasons. For obvious reasons it is critically important to have an accurate estimate of the time at peak magnification, which influences many other light curve fitting parameters like $u_0$, $t_E$, $f_S$, $\pi_E$, etc. Similar to the CROIN parameterization, our next cutting threshold is on the value of the impact parameter, $u_0$, also referred to as the lens-source closest approach. We restrict the absolute value of $u_0$ to be less than 3 in order to cause a magnification signal from the primary that is detectable at the expected \textit{Roman} photometric precision (see Table \ref{tab:survey-params}). We then make a cut on the catalog for events that fall outside of the 5-field `Overguide' GBTDS footprint. Note that we allow a small padding, mostly in galactic longitude ($\ell$) space, due to the 180 degree rotation in the WFI orientation between northern spring and autumn seasons.\\
\indent At this point our sample consists of approximately 61,500 simulated events across the six planet mass bins. We estimate that processing this number of events through \texttt{pyLIMASS} would take approximately 10 months on a 16-core CPU, 64GB-memory machine. Given this number of events is ${\sim} 1.6$ orders of magnitude larger than the expected yield of microlensing planets from \textit{Roman} \citep{penny19, zohrabi:inprep}, we decide to implement one final culling of the dataset. The bottom row of Table \ref{tab:cutting-thresholds} describes our weighted sampling of the remaining events. For each event $i$ in each of the six mass bins $M_{p,b}$, each of which has $N_b$ events per bin, we assign a probability $p_i,b$ that is proportional to the normalized weight $w_i$ of the event as described in Section 3. We define the normalized probability of event $i$ in mass bin $b$ to be $p_{i,b} = \dfrac{w_i}{\sum_{b=1}^{N_b} w_i}$,  where the sum is over the $N_b$ events in mass bin $b$.  We then sample 500 events from each mass bin according to their individual probability $p_{i,b}$. This results in our final selection of 3,000 total events across all six mass bins. Note in our final selection, we remove duplicate events (e.g. sampling with replacement), so each event in the final sample is unique. The right panel of Figure \ref{fig:sim_events_figure} shows the 3,000 simulated events we analyze in this work, colored by $K_S-$band extinction. The extinction models used in the simulations come from \texttt{SynthPop} and are described in more detail in Section \ref{subsec:galactic_models}. \\
\indent An important detail to consider for our simulated sample of events is how the associated errors for various microlensing-related parameters are derived. We give a brief description here and refer the reader to \cite{zohrabi:inprep} for a full discussion. The methodology that is adopted is the Fisher information matrix (FIM); the covariance matrix of the best unbiased estimator is given by the inverse of the FIM. The diagonal terms in this matrix give the variance of the parameter estimates, and the off-diagonal terms describe the correlation between discrete parameters. Fisher matrices are an efficient way of estimating errors on large multi-dimensional datasets that would otherwise take significant time and resources to generate with other methods like Bayesian inference or MCMC sampling \citep{rodriguez:2013a}. Figure \ref{fig:uncertainty_grid} shows the distributions of parameter errors for the observables that we process through \texttt{pyLIMASS}. We note the parameters in Figure \ref{fig:uncertainty_grid} for which Fisher errors are derived are $t_0, t_E,\rho,\pi_{EE},$\, and $\pi_{EN}$ (grey histograms). The distributions of errors on $\mu_{\textrm{rel}}$, F$_{\textrm{source}}$, and F$_{\textrm{lens}}$ (colored histograms) are derived using an analytic form based on the image elongation or CDCS measurement. Further details on the derivation of this analytic form can be found in Sections \ref{subsec:lens-flux-description}, \ref{subsec:murel_description}, and \cite{verma:inprep}. \\
\indent We note that the requirement we are validating is on the host mass (and distance), not the planet mass.  However, the planet mass $M_p$ is arguably of equal, or perhaps even more, interest than the host mass. Thus, we might want to consider the uncertainty on $M_p$, which depends on the uncertainty in the measurements of $M_L$ and the mass ratio $q$, as well as any covariances between these measurements. However, as \texttt{pyLIMASS} does not consider $q$ or its uncertainty, we cannot use the same methodology we use to derive the uncertainties on the other parameters we consider to estimate the uncertainty on $M_p$. As \cite{bachelet:2024a} describe, the primary function of \texttt{pyLIMASS} is to estimate the physical properties of the lensing systems, including the direct mass of the lens in units of $M_\odot$.  We could simply quote the uncertainty on $M_p$ as the quadrature sum of the uncertainties on $q$ (derived from the Fisher matrix estimate) and $M_L$ derived from \texttt{pyLIMASS}, but this would ignore the observational covariances between these parameters, which are generally expected to be small but may not be negligible. Furthermore, as we alluded to earlier in Section \ref{subsec:high-order-effects}, this procedure would implicitly assume that the probability of a given star hosting a planet of a given mass ratio is independent of $M_L$, which is unlikely to be correct in practice \citep{bennett:inprep}.  We therefore choose to focus on the uncertainty in $M_L$ (and $D_L$).

\subsection{White Dwarf \& Very Low Mass Stellar Hosts} \label{sec:bdwds}
\begin{figure}[!h]
\includegraphics[width=1.0\linewidth]{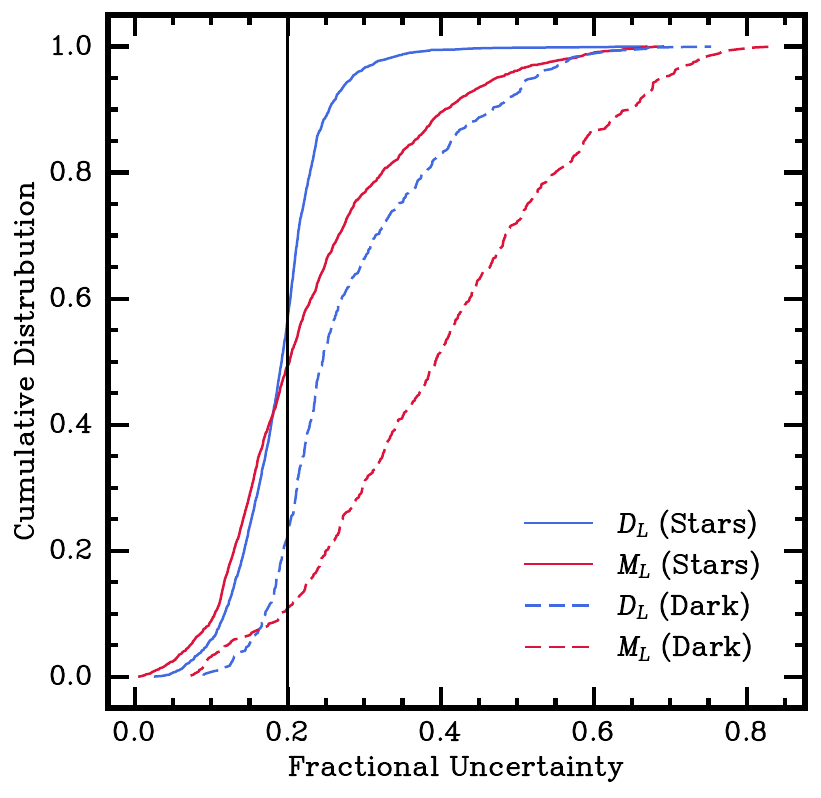}
\centering
\caption{\footnotesize Cumulative Distributions for the lens mass and distance estimate from \texttt{pyLIMASS}. The solid vertical line indicates the 20\% fractional error requirement. Solid curves are the $M_L$ and $D_L$ distributions for stars, dashed curves are the distributions for dark lenses (WDs, VLMs). \label{fig:CDF}}
\end{figure}

We expect ${\sim}20$\% of all lens hosts to be comprised of white dwarfs (WDs) or brown dwarfs (BDs) \citep{gould:2000a, bennett:2007a, gaudi:2012a}, with a $\la 1\%$ contribution from neutron stars and black holes, which we ignore here. Along with WD and BDs, very low mass (VLM) at the bottom of the main sequence are also expected to make up a small fraction of the lens hosts, these objects are also very dim. We note the isochrones used to generate the stellar population in \texttt{GULLS} have a lower-limit of $0.1M_{\odot}$, which is $0.02M_{\odot}$ above the lower mass limit of the stellar population that has been generated via \texttt{SynthPop}. Therefore we denote the range of host masses between $0.08-0.10M_{\odot}$ as VLMs. These hosts do not have a measurable brightness in any \textit{Roman} passband, and we group them into the dark category along with WD hosts. Currently \texttt{GULLS} does not support BDs or BD isochrones, so our subsequent analysis does not include planet-hosting BDs.\\
\indent As a check on our sampled population, we confirm that WDs and VLMs make up an appropriate proportion (in our case ${\sim}$19\%) of the 3,000 events we've selected for processing with \texttt{pyLIMASS}. Figures \ref{fig:CDF}, \ref{fig:individ_events_hist}, and \ref{fig:sim_events_correlations} distinguish the sample of stellar hosts (e.g. luminous lenses) from the sample of dark hosts (non-luminous lenses). Since all of these dark objects do not have appreciable flux in \textit{Roman}, a precise determination of the lens mass and distance must rely on both higher-order effects to be well-measured from the simulated light curve photometry, as previously explained in Section \ref{subsec:high-order-effects}. Lastly, it may be possible to estimate $thete_{\rm E}$ via precise astrometry of the source, whose centroid shifts during the microlensing event due to the time-varying fluxes and positions of the unresolved micro-images created during the event. This signature would be largest for the most massive and closest host lenses, and may offer an avenue to improve the fraction of events with constraints on the lens masses and distances \citep{dominik:2000a,lam:2023a,sajadian:2023a}. This is a topic of active investigation within the RGES PIT, however it is beyond the scope of work in our current study.

%------------------pyLIMASS---------------------------------
%------------------------------------------------

\section{\texttt{pyLIMASS}} \label{sec:pylimass}
First introduced by \cite{bachelet:2024a}, \texttt{pyLIMASS} is an additional software module incorporated into the python Lightcurve Identification and Microlensing Analysis \texttt{pyLIMA} suite \citep{bachelet:2017a}. The primary function of \texttt{pyLIMASS} is to compute the lens system physical properties (e.g. mass, distance) using all available information regarding the microlensing event. The primary inputs to \texttt{pyLIMASS} are called `observables' and they can include light curve parameters (e.g. $t_E,\, t_0,\, \rho,\, \pi_E$, etc), high-resolution follow up information (e.g. lens flux), and even completely external information (e.g. \textit{Gaia} parallax). \\
\indent The code relies on PARSEC stellar isochrone models \citep{bressan:2012a}, and combines all input observables into a multidimensional Gaussian mixture (GM) model (e.g. \cite{fruhwirth:2019a}). One advantage of the GM approach is its simplicity and efficiency with which complex, multi-modal distributions can be evaluated. \cite{bachelet:2024a} have shown that their implementation of the GM can successfully approximate the underlying probability distribution of a sample of $N$ observables. For a full description of the GM and the overall \texttt{pyLIMASS} structure, we refer the reader to \cite{bachelet:2024a}.\\
\indent \cite{bachelet:2024a} simulated 1000 PSPL microlensing events observed in \textit{Roman} passbands F087 and F146, including details of the observing strategy given in \cite{penny19} (50 sec exposure time, zero-point magnitude of $Z_P=27.4$ mag). The observables that were used for these simulated events are: $t_E, F087_S, F146_S, A_{V_S}, F087_L, F146_L, \theta_*, \pi_{\textrm{EN}}, \pi_{\textrm{EE}}, \mu_{\textrm{rel}}$. Figure 3 of \cite{bachelet:2024a} shows the distributions of the accuracies with which the lens masses and distances of these simulated events are reconstructed, specifically the distributions of $\delta M_L$ and $\Delta D_L$, the difference between the values of $M_L$ and $D_L$ inferred from \texttt{pyLIMASS} and their true values. A correlation can also be seen in their distribution of source-lens brightness difference ($\Delta F146$) as a function of lens mass. This is because $\theta_E$ can be measured from the ingested observables ($t_E$ and $\mu_{\textrm{rel}})$, and the lens mass can be approximated by $M_L\,{\approx}\, \theta_{E}^2 D_L/\kappa$, where $\kappa = 4G/c^2$. They demonstrated that the medians of the $\delta M_L$ and $\Delta D_L$ values are close to zero, although they do find a bias in $M_L$ for events where the source and/or lens brightness are poorly measured.\\

\subsection{Estimating Masses \& Distances} \label{subsec:ml-dl-pylimass}

\indent The initial \cite{bachelet:2024a} results demonstrate that \texttt{pyLIMASS} can deliver lens masses with a median precision of $\sigma_{M_L}/{M_L} = 20\%$, with little to no bias. There is no conclusion given for the median precision of lens distances $\sigma_{D_L}/{D_L}$ from their simulations, and thus they cannot be used to validate the distance component of the requirement. \\
\indent  In general, \cite{bachelet:2024a} find that the masses and distances to the sources are less precisely determined than those of the lenses. This is not surprising, as the only observables that constrain the physical properties of the source are the source flux and color. These are related to the mass and distance to the source through the isochrones used by \texttt{pyLIMASS}. However, \texttt{pyLIMASS} does not implement any prior on physical parameters of the source (or lens) based on, e.g., a galactic model of the spatial and kinematic distribution of stars. Thus, the source distance (as well as the source mass) is essentially a free parameter, and even assuming precisely-measured extinction, the source mass and distance cannot be uniquely determined from the source flux and color alone. As a result, \texttt{pyLIMASS} has a tendency to overestimate the source distance, which in turn results in overestimated lens distances. Including galactic priors can improve the constraints on the the source distance (and mass) and thus on the lens mass and distance. \\
\indent Lastly, we note that \cite{bachelet:2024a} performed a \texttt{pyLIMASS} analysis on five published planetary microlensing events and one published stellar binary microlensing event. A majority of these events have high-resolution followup imaging data to constrain the lens brightness and lens-source relative proper motion, similar to the historical events analyzed in this work (Section \ref{sec:modeling-pylimass}). Table 2 of \cite{bachelet:2024a} reports the \texttt{pyLIMASS} estimates and uncertainties for lens mass, lens distance, and source distance for their six-event sample. Generally, the \texttt{pyLIMASS} ($M_L, D_L$) uncertainties reported for their targets are consistent with the uncertainties we find in this work (e.g. Figure \ref{fig:pylimass-vs-empirical} and \texttt{pyLIMASS} rows in Table \ref{tab:eesunhong-pylimass}).

\subsubsection{Source Distance Prior via Galactic Models}\label{subsec:galactic_models}
Recall that a measurement of the distance to the lens through measurements of $\theta_E$ and $\pi_E$ depends on the distance to the source via Equation \ref{eqn:dl-sl}. 

\begin{figure}[!h]
\includegraphics[width=1\linewidth]{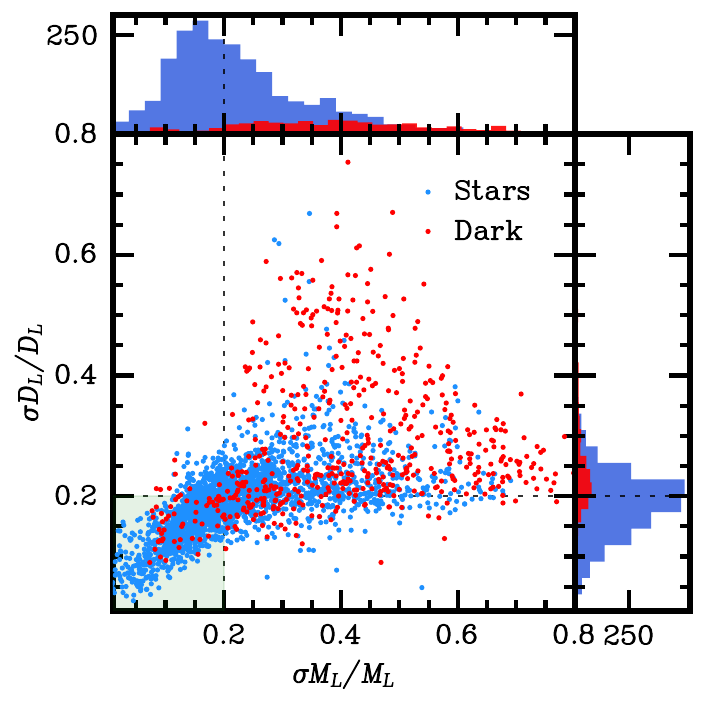}
\centering
\caption{\footnotesize Distribution of $M_L$ and $D_L$ fractional errors for 3,000 simulated events processed by \texttt{pyLIMASS}. Stars are shown as blue points and histograms, and dark lenses (WDs, VLMs) are shown as red points and histograms. \label{fig:individ_events_hist}}
\end{figure}

\noindent Therefore, we aim to generate a well understood distribution of source distances for the sample of simulated microlensing events that we evaluate with \texttt{pyLIMASS}. To this end, we establish a prior probability distribution on the source distance for each event (and sight-line toward the bulge). We have implemented two different galactic models in \texttt{pyLIMASS} to generate these distributions; the galactic model of \cite{koshimoto:2021a} (hereafter KM21) and the \texttt{SynthPop}-based galactic model of \cite{huston:inprep}, hereafter SP-H24. \\
\indent KM21 is a parametric galactic model built using constraints from the spatial distribution of velocity and velocity dispersions in \textit{Gaia} DR2 \citep{gaia:2018a}, OGLE-III red clump star counts \citep{nataf:2013a}, VIRAC proper motion data \citep{smith:2018a}, BRAVA radial velocity data \citep{rich:2007a}, and OGLE-IV star count and microlensing event rates \citep{mroz:2017nature, mroz:2019a}. The purely parametric nature of the model allows for easy implementation and reproducibility. Nine of the 15 historical events we've selected in this work have used the accompanying code \texttt{genulens} \citep{koshimoto:2021code} to generate the prior probability distributions for source distances using the sky coordinates for the given event. The remaining six historical events used a series of older galactic models that were developed before KM21 (see Section \ref{sec:modeling-pylimass}). \\

\indent \texttt{SynthPop} is a flexible Galactic modeling tool where various components of the Milky Way can be easily modified and tuned (\citetalias{kluter:2025a}, \citeyear{kluter:2025a}). The implementation of SP-H24 is a preliminary model version used for Roman microlensing simulations. The framework uses a Galactic bulge density model, which includes the thin and thick disks, and the stellar halo. These bulge and disk components are based on microlensing survey results, and the halo is relatively unimportant in terms of microlensing survey statistics, particularly in the direction of the Galactic bulge and plane. The bulge and disk kinematic models are pulled from \cite{koshimoto:2021a} and halo kinematics from \cite{Robin2003}. For all populations, the model adopts the \citet{Kroupa2001} initial mass function and MIST \citep[MESA Isochrones and Stellar Tracks]{Choi2016,Dotter2016} isochrones. For compact objects beyond the MIST grids, it uses the \texttt{SukhboldN20} scheme from \texttt{PopSyCLE} \citep{lam:2020a}, described in Sec. 3.4 of \citealp{Rose2022}, which is subsequently adopted from \cite{Sukhbold2014,Sukhbold2016, Woosley2017, Woosley2020} and \cite{Kalirai2008}. For extinction, the model uses the 2-d map of \citet{Surot2020}, with an extinction law from \citet{Cardelli1989}; the optical side is set by \citet{O'Donnell1994} and infrared side is adjusted to match \citet{Surot2020}. \cite{huston:inprep} will present this model fully, with evaluation against data, as well as the updated SP-H25 model which adds a nuclear stellar disk population and alters the extinction model to be 3-dimensional. Since the SP-H24 model was used for the \texttt{GULLS} event simulation, we use this model to generate the source distance prior probability distributions for our sample of 3,000 simulated events. \\
\indent A subtle but important caveat worth mentioning is that there is a difference in the isochrones used to generate the stellar population in \texttt{SynthPop} (MIST), and the isochrones that \texttt{pyLIMASS} uses for its mass-luminosity relations (PARSEC). This means that our analysis is effectively probing the systematic uncertainty due to the different choices of isochrones. Therefore, the magnitude of the differences between the isochrones represents the systematic uncertainties in our ability to model the true mass-luminosity relationship of the stars of interest. It would appear that this systematic uncertainty in our inferred parameters due to this source of astrophysical uncertainty is relatively small given our results presented in Section \ref{subsec:pylimass_results}.\\
\indent Lastly, we have performed a comparison using the KM21 and SP-H24 models to establish the prior probability distribution on source distances for 5\% of our total sample. We ensure the sub-sample we select for this comparison spans the nearly ${\sim}1\degree$ in galactic latitude that is covered by the GBTDS footprint (shown in Figure \ref{fig:sim_events_figure}). We fix all observables and only vary the choice of galactic model for each event, and we find at most a ${\sim}1\%$ difference in the mean value of source distance ($\overline D_S$) and dispersion ($\sigma_{D_S}$) between the models. Since the dependence of $D_L$ on $D_S$ is non-linear (equation \ref{eq:dl-ls}), assessing the impact of this 1\% difference requires knowledge of the magnitude of $\pi_E \theta_E$. At most we may expect a ${\sim}2\%$ difference in the inferred lens distance $D_L$ between KM21 and SP-H24. For the majority of our simulated lenses at ${\sim}7.5$ kpc (Figure \ref{fig:sim_events_correlations}) this translates to a relatively minor difference of 0.15 kpc.

\subsection{\texttt{pyLIMASS} Results}\label{subsec:pylimass_results}

Finally, we use \texttt{pyLIMASS} to determine lens masses, distances, and associated uncertainties for all 3000 simulated events in our final sample. For the input observables, we have included light curve fitting parameters with associated fisher uncertainties from \cite{zohrabi:inprep} (Figure \ref{fig:uncertainty_grid}), as well as estimates of the lens flux (and errors) and estimates of lens-source relative proper motion (and errors) via the image elongation and color-dependent centroid shift methods \citep{verma:inprep} (Section \ref{subsec:murel_description}). For the ${\sim}20\%$ of simulated events with dark lens hosts, there is no measurement of lens flux or lens-source relative proper motion, so those observables have been omitted.
\begin{figure*}[!htb]
\includegraphics[width=0.9\linewidth]{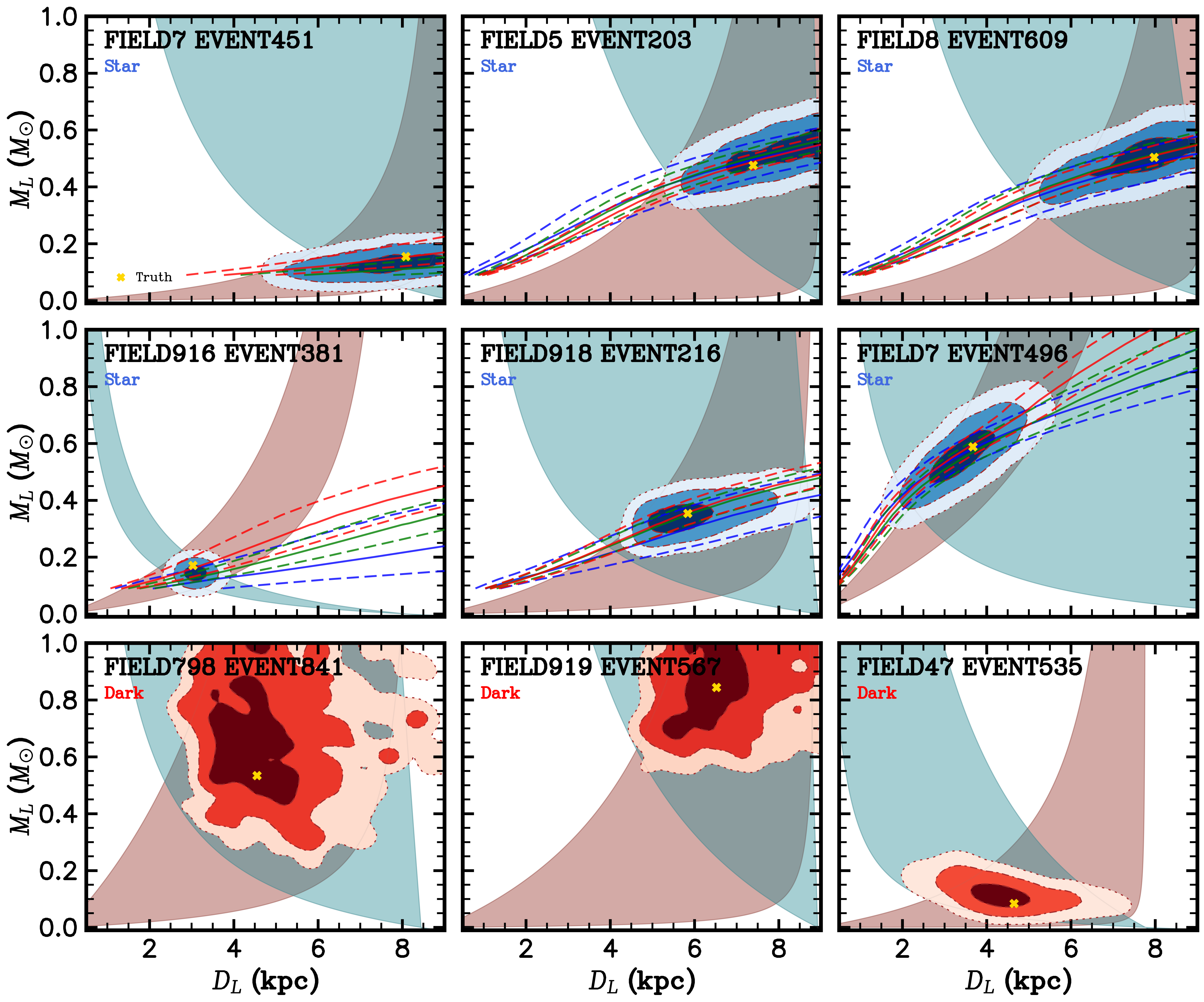}
\centering
\caption{Estimated masses and distances for a representative sample of simulated events. The constraints from an estimate of $\theta_E$ are shown as brown shaded curves, and the $\pi_E$ constraints are shown as teal shaded curves. The blue, green, and red lines show the constraint from the measured flux in the \textit{Roman} $F087$, $F146$, and $F213$ passbands, combined with mass-luminosity relations from the \texttt{pyLIMASS}-implemented PARSEC isochrones \citep{bressan:2012a}. The dashed lines show the $1\sigma$ errors that come from the quadrature sum of the lens flux uncertainties and an assumed 5\% intrinsic uncertainty on the mass-luminosity relations. The yellow data points show the true value of the lens mass and distance. The blue contours give the \texttt{pyLIMASS} posterior distributions (central 39\%, 86\%, 99\%) for stellar (luminous) lenses, red contours (bottom row) give the posteriors for three dark lenses (two WDs and one VLM star). \label{fig:MD_grid_simEvents}}
\end{figure*}
\indent Figure \ref{fig:CDF} shows the cumulative distribution for all simulated events as a function of the fractional error delivered by \texttt{pyLIMASS}. The curves are broken into stars (e.g. luminous lens hosts) as solid lines, and dark (dim or non-luminous) as dashed lines. The estimate of lens distance ($D_L$) are given in blue, and lens mass ($M_L$) are given in red. As expected, the sample of luminous stellar hosts has a significantly higher fraction with $D_L$ and $M_L$ estimates at or better than 20\%, compared to the dark lens hosts. This is not surprising, as we can only measure at most two of the three possible mass-distance relations mentioned earlier in Sections \ref{subsec:high-order-effects} and \ref{subsec:lens-flux-description}.\\

\begin{figure*}[!htb]
\includegraphics[width=0.9\linewidth]{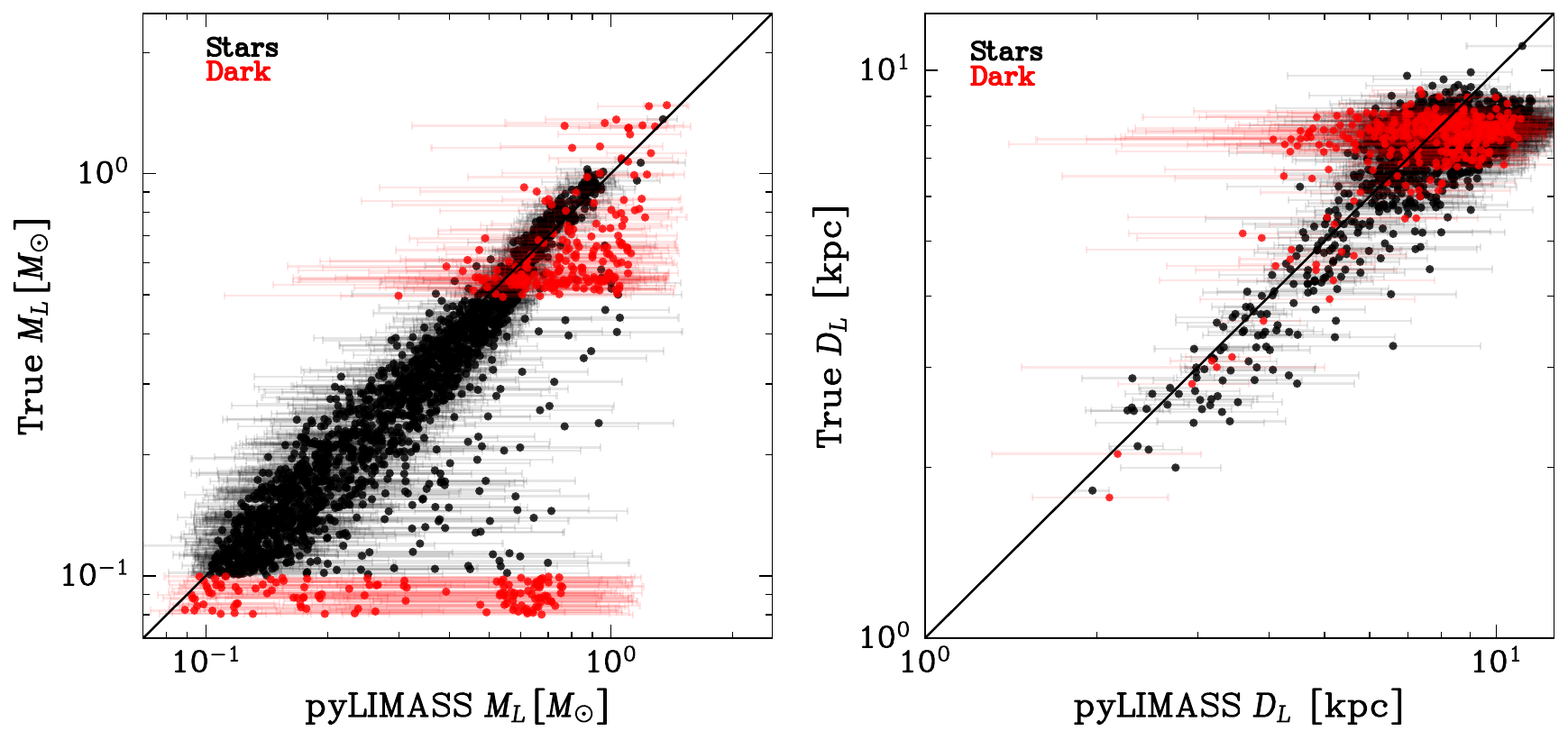}
\centering
\caption{Simulated events True vs. \texttt{pyLIMASS}-measured values of lens mass (\textit{Left}) and lens distance (\textit{Right}). The solid line denotes the 1:1 correlation. To improve readability, we remove $> 3\sigma$ outliers (${\sim}6\%$ of events) from both panels. \label{fig:sim_events_correlations}}
\end{figure*}

\indent Figure \ref{fig:individ_events_hist} shows the fractional errors in $D_L$ and $M_L$ for each of the simulated events. Again the (luminous) `Star' host lenses are shown as the blue data points and histograms, and the (non-luminous) dark host lenses are shown as red data points and histograms. The dashed lines denote the 20\% fractional error threshold governed by the science requirement. Thus the lenses that lie inside the lower-left green shaded region in Figure \ref{fig:individ_events_hist} represent events that satisfy the requirement (e.g. both $D_L$ and $M_L$ estimates have 20\% precision or better). Lastly, we note a trend can be observed in the Figure \ref{fig:individ_events_hist} distributions, particularly in the luminous stellar lens population (blue data points). There is a noticeable break in the distribution, where at $\sigma M_L /M_L < 0.20$ the distribution is roughly linear. However, beyond $\sigma M_L /M_L > 0.20$ the distribution flattens along a vector aligned with $\sigma D_L /D_L \sim 0.20$. We attribute this trend to the source distance ($D_S$) prior probability distribution generated using the \texttt{SynthPop} galactic model for each simulated event. As can be seen in equation \ref{eq:dl-ls} and Section \ref{subsec:galactic_models}, the choice of $D_S$ can have a direct influence on the resulting $D_L$ distribution. \\
\indent Regardless of whether or not a $D_S$ prior from a galactic model is used, \textit{Roman} sensitivity on real data may allow for trigonometric parallax measurements for source stars if the zero point of astrometric grids can be properly calibrated. This will mitigate the need to use a galactic model to generate $D_S$ prior probabilities for real \textit{Roman} microlensing events. For instance, if we assume the fractional uncertainty in the source parallax from photon noise is $\sigma_{\pi_S} / \pi_S\, {\sim}\, (\sigma_x / \sqrt{N_d/2}) {\pi_S}^{-1}$ where $\sigma_x$ is the per-epoch uncertainty on the source position, $N_d\, {\sim}\, 430$ days/(12 minutes) ${\sim}$ 51,600, and $\pi_S\, {\sim}\, 0.125$ mas ($D_S$/8 kpc)$^{-1}$ is the parallax of the source. For an isolated star with $F146\, {\sim}\, 22$, we can expect that $\sigma_x\, {\sim}\, 1$ mas, so we can expect that $\sigma_{\pi_S} / \pi_S\, {\sim}\, 5\%$. Of course the situation is made more complicated by luminous lenses highly blended with the sources, particularly lenses that have brightnesses of order or brighter than the source. \\
\indent We have selected a sub-sample of nine simulated events to display in Figure \ref{fig:MD_grid_simEvents}. The events are evenly distributed across the GBTDS footprint and we have sampled at least one event from each planet mass bin. Each panel shows a wide lens mass and distance parameter space ($0.01M_{\odot} < M_L < 1.0M_{\odot}$ and $0.5$\,kpc $< D_L < 9$\,kpc). The top two rows of Figure \ref{fig:MD_grid_simEvents} show stellar (luminous) lenses, and the bottom row shows dark lenses. The combination of lens flux detections and at least one of two light curve higher-order effects will enable robust lens mass and distance estimates during the GBTDS. In particular, this figure demonstrates the importance of lens flux measurements in (most) cases where the microlensing parallax is not well-measured from the light curve data. \\
\indent Figure \ref{fig:sim_events_correlations} shows the comparisons of $M_L$ (\textit{left} panel) and $D_L$ (\textit{right} panel) for the true values relative to the \texttt{pyLIMASS}-estimated values (and uncertainties) for the simulated events in the sample. The inferred values of $M_L$ and $D_L$ for the stellar (luminous) population agree well with the true values. The dark population of lenses are in poorer agreement, for the reasons discussed earlier. We note that there is a peak in the distance distribution of lenses near $D_L \sim 7.5~{\rm kpc}$, implying that the majority of exoplanet systems that Roman will detect may be in the bulge, although of course this depends on how the frequency of exoplanets depends on host star properties and environment. For clarity in Figure \ref{fig:sim_events_correlations} we remove outliers with \texttt{pyLIMASS}-estimated $M_L > 3\sigma$ from the true $M_L$ value, this accounts for ${\sim}6$\% of simulated events. This population of outliers may be due to expected statistical fluctuations, subtle biases inherent in \texttt{pyLIMASS}, true degeneracies in some of the simulated events, or likely a combination of these factors. We note that less than $0.5\%$ of simulated events are $> 5\sigma$ outliers in $M_L$. \\
\indent Lastly, Table \ref{tab:pylimass_requirement} gives the overall results for the fraction ($f$) of simulated events with $M_L-$only estimates, $D_L-$only estimates, and joint $M_L$ \& $D_L$ estimates at 20\% precision or better. We assess the impact of any intrinsic variance in the \texttt{pyLIMASS} inferences by repeating running \texttt{pyLIMASS} five times for each mass bin, and then computing the mean and standard deviations in the inferred physical parameters over the five trials. For each row in Table \ref{tab:pylimass_requirement} we report the mean value of the five repeat runs. The `Total' in Table \ref{tab:pylimass_requirement} shows the total fraction integrated across each row, scaled by the \cite{cassan:2012a} mass function (MF) as described earlier in Section \ref{sec:requirement}. There is a significant anti-correlation between the fraction of events for which $M_L$ and/or $D_L$ is measured to a precision of 20\% and the planet mass. As the planet mass ($M_{\oplus}$) increases, the fraction of events ($f$) with 20\% or better precision decreases. This is primarily due to the detectability of finite source effects in the simulated light curves. Lower mass planets give a preferentially higher sensitivity to finite source effects, particularly in high magnification events, as compared to events with higher mass planets. This downward trend continues until we reach the largest planet mass bin ($10^4M_{\oplus}$), which shows a rise to higher efficiencies once more. This upturn is largely due to the fact that lenses with higher-mass planetary companions have preferentially higher sensitivity to microlensing parallax detections, which leads to higher precision on the $M_L$ and $D_L$ estimates. \\

\begin{deluxetable*}{@{\extracolsep{4pt}}lccccccccc}[!htb]
\tablecaption{Lens Mass \& Distance Uncertainty Results \label{tab:pylimass_requirement}}
\setlength{\tabcolsep}{0.1pt}
\tablewidth{\columnwidth}
\tablehead
{
\colhead{}&
  \multicolumn{3}{c}{$M_L$ only}&
  \multicolumn{3}{c}{$D_L$ only}&
  \multicolumn{3}{c}{$M_L$ \& $D_L$}\\
\cline{2-4} \cline{5-7} \cline{8-10}
\colhead{\hspace{-0.01cm}Planet Mass ($M_{\oplus}$)} & \colhead{$f_{\textrm{Stars}}$}& 
\colhead{$f_{\textrm{Dark}}$} & \colhead{$f_{\textrm{All}}$} & \colhead{$f_{\textrm{Stars}}$} & \colhead{$f_{\textrm{Dark}}$} & \colhead{$f_{\textrm{All}}$} & \colhead{$f_{\textrm{Stars}}$} & \colhead{$f_{\textrm{Dark}}$} & \colhead{$f_{\textrm{All}}$}
}
\startdata
0.1 & 59.7\% & 18.7\% & 52.0\% & 68.6\% & 43.7\% & 64.0\% & 54.0\% & 15.2\% & 46.8\%\\
1 & 53.2\% & 10.3\% & 45.7\% & 63.2\% & 25.0\% & 56.5\% & 45.0\% & 7.3\% & 38.4\%\\
10 & 53.0\% & 6.9\% & 43.0\% & 57.8\% & 16.9\% & 49.0\% & 42.5\% & 5.3\% & 34.5\%\\
100 & 44.3\% & 8.6\% & 37.8\% & 55.7\% & 18.9\% & 49.0\% & 36.6\% & 8.1\% & 31.4\%\\
1000 & 41.9\% & 5.1\% & 35.0\% & 48.6\% & 11.9\% & 41.7\% & 33.0\% & 4.5\% & 27.7\%\\
10000 & 44.7\% & 15.2\% & 38.8\% & 50.0\% & 20.0\% & 44.0\% & 35.3\% & 11.1\% & 30.5\%\\
\hline
%Raw Total ($0.1-10$$^4$ $M_{\oplus}$) & {49.2}\% & {10.5}\% & {41.6}\% & {56.5}\% & {22.1}\% & {49.8}\% & 40.9\% & 8.9\% & 34.7\%\\
%\hline
Total & $55.0 \pm 0.6\%$ & $12.5 \pm 1.0\%$ & $47.0 \pm 0.5\%$ & $63.5 \pm 0.5\%$ & $29.6 \pm 1.3\%$ & $57.1 \pm 0.5\%$ & $47.2 \pm 0.4\%$ & $9.7 \pm 0.8\%$ & $40.2 \pm 0.5\%$\\
\enddata
\tablenotetext{}{\footnotesize{\textbf{Note}. $f$ represents the (mean) fraction of simulated events that have lens mass and/or distance measurements with 20\% precision or better. The last row with total values has been scaled by the \cite{cassan:2012a} mass function, as described in Section \ref{subsec:pylimass_results}.}}
\end{deluxetable*}

%---------------------------------------Image-Constrained Modeling-----------------------------------------------------------------------------------------------------------------------------------------

\section{Image-constrained Light Curve Modeling and \texttt{pyLIMASS}} \label{sec:modeling-pylimass}
As we previously mention in Section \ref{subsec:lens-flux-description}, directly measuring the lens (host) flux can be very powerful for constraining the lens system physical properties. Successful implementations of this method have been demonstrated using high-resolution follow up imaging campaigns for over 10 years now. The technique of incorporating high-resolution imaging results into the light curve modeling, referred to as \textit{image-constrained light curve modeling}, was formally introduced by \cite{bennett:2024a}. This is a forward modeling approach where constraints from high angular resolution imaging of the microlensing target are applied at the light curve modeling level. Many plausible models that would normally be allowed by light curve photometry-only can be excluded if the high angular resolution imaging data strongly disfavors them. Detailed descriptions of this method can be found in \cite{bennett:2024a}, and a public code that supports image-constrained light curve modeling is \texttt{eesunhong}\footnote{\url{https://github.com/golmschenk/eesunhong}}. 

\begin{figure}[!h]
\includegraphics[width=1\linewidth]{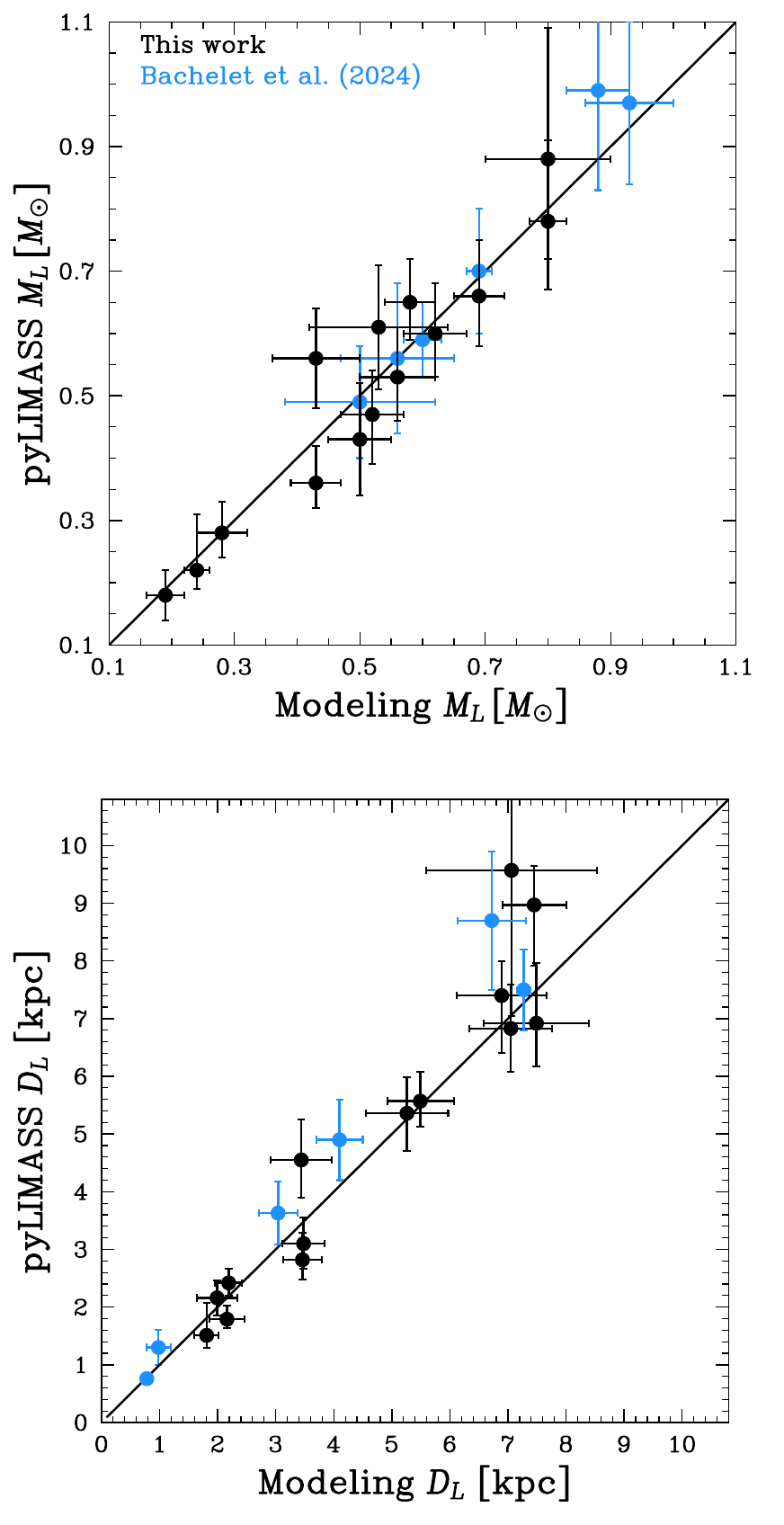}
\centering
\caption{\footnotesize Comparison between \texttt{pyLIMASS}-estimated lens mass and distance for 15 historical events evaluated in our work (black data points). We include the six events analyzed by \cite{bachelet:2024a} in blue. \label{fig:pylimass-vs-empirical}}
\end{figure}

\indent One of our motivations here is to compare the lens mass and distance values that are independently derived by image-constrained modeling (\texttt{eesunhong}) and \texttt{pyLIMASS}. We are essentially comparing two conceptually different frameworks for estimating posteriors on the physical parameters. In the case of \texttt{eesunhong}, we are directly sampling the likelihood surface of the parameters by evaluating the likelihood function at different points in parameter space, and then using MCMC to determine the posteriors. In the case of \texttt{pyLIMASS}, we are combining posterior distributions of the observable parameters to determine the values of the physical parameters that best match these posterior distributions, assuming a Gaussian mixture model. We expect the RGES survey will take advantage of both techniques to analyze planetary microlensing events during the GBTDS, therefore it is necessary to determine if there exists any biases in either method.\\
\indent For our empirical study, we select a sample of 15 planetary microlensing events, all of which have high angular resolution follow-up imaging with Keck AO, \textit{HST}, or both Keck AO $+$ \textit{HST}. As previously described in Section \ref{subsec:galactic_models}, the choice of source distance prior via galactic models can influence the inferred lens distance. The discovery papers for historical events we analyze have used various methods and galactic models to derive prior distributions on source distances \citep{han:2003a, gould:1994a, robin:2003a, dominik:2006a, bennett:2014a, mroz:2020a, koshimoto:2021a}. While this is a necessary caveat to consider, a full re-analysis of each historical microlensing event using an identical galactic model is beyond the scope of our current work. \\
\indent Table \ref{tab:eesunhong-pylimass} lists the 15 planetary events, the discovery paper references, as well as the high-resolution follow up analysis references. We include lens mass and distance, and source distance values and one sigma errors reported from modeling of only the light curve photometry and from image-constrained light curve modeling. We run \texttt{pyLIMASS} on both scenarios and report the corresponding physical parameters and one sigma errors for comparison. In order to avoid double-constraining certain parameters in the \texttt{pyLIMASS} run that includes constraints from high-resolution imaging, we keep the light curve-only related parameters (and errors) consistent between both trials. For example, if the measurement precision on $\pi_E$ improves as a result of a direct measurement of ($\mu_{\textrm{rel,N}}, \mu_{\textrm{rel,E}}$) via high-resolution imaging, we ignore this in \texttt{pyLIMASS} and use the light curve-only value (and error) on $\pi_E$. This means that the only new information we give \texttt{pyLIMASS} in the ``Light curve$+$Imaging" trial (e.g. right-hand columns of Table \ref{tab:eesunhong-pylimass}) is the direct lens flux measurement and error in each available passband, as well as the lens-source relative proper motion ($\mu_{\textrm{rel}})$ that comes from a direct measurement of the lens-source separation. Further, Table \ref{tab:observables-priors} gives a list of all \texttt{pyLIMASS} microlensing observables and priors that were used for each of the 15 historical events we analyze. As mentioned previously, we invoke a source distance prior in the form of a normal distribution for all 15 events. We note \texttt{pyLIMASS} does not currently support WD isochrones, therefore we are required to include an additional host mass prior for the one WD lens in the sample, MOA-2010-BLG-477 \citep{blackman:2021a}. This prior is in the form of a normal distribution centered at 0.6$M_{\odot}$ with $\sigma = 0.1M_{\odot}$. \\
\indent For this empirical analysis that includes high-resolution follow up imaging, there are some important caveats to consider:

\begin{enumerate}
    \item A typical set of ${\sim}16$ \textit{HST} (or \textit{Keck}) images from historical high-resolution campaigns is ${\sim}2.5$ orders of magnitude less than the expected number of images that will come from \textit{Roman} (Table \ref{tab:survey-params}).
    \item RGES photometry and astrometry will come from the same dataset, so we expect to have better constraints on the source fluxes than we have historically had with estimates from ground-based observations.
    \item Most of the RGES-measured lens-source separations will be smaller than the historical \textit{Keck/HST} sample, and the \textit{Roman} WFI pixel scale is larger than \textit{HST}-UVIS and \textit{Keck}. On the other hand, the PSFs will be measured with significantly higher SNR in the GBTDS, and the precise shape of the PSF will be know to exquisite precision. Furthermore, the primary filter is in the near-infrared, where the difference in magnitude between the sources and lenses will generally be smaller than in bluer \textit{HST} observations considered here for most of this sample, particularly for the \textit{HST} $V-$band observations.
\end{enumerate}

\begin{figure*}[!htb]
\includegraphics[width=0.9\linewidth]{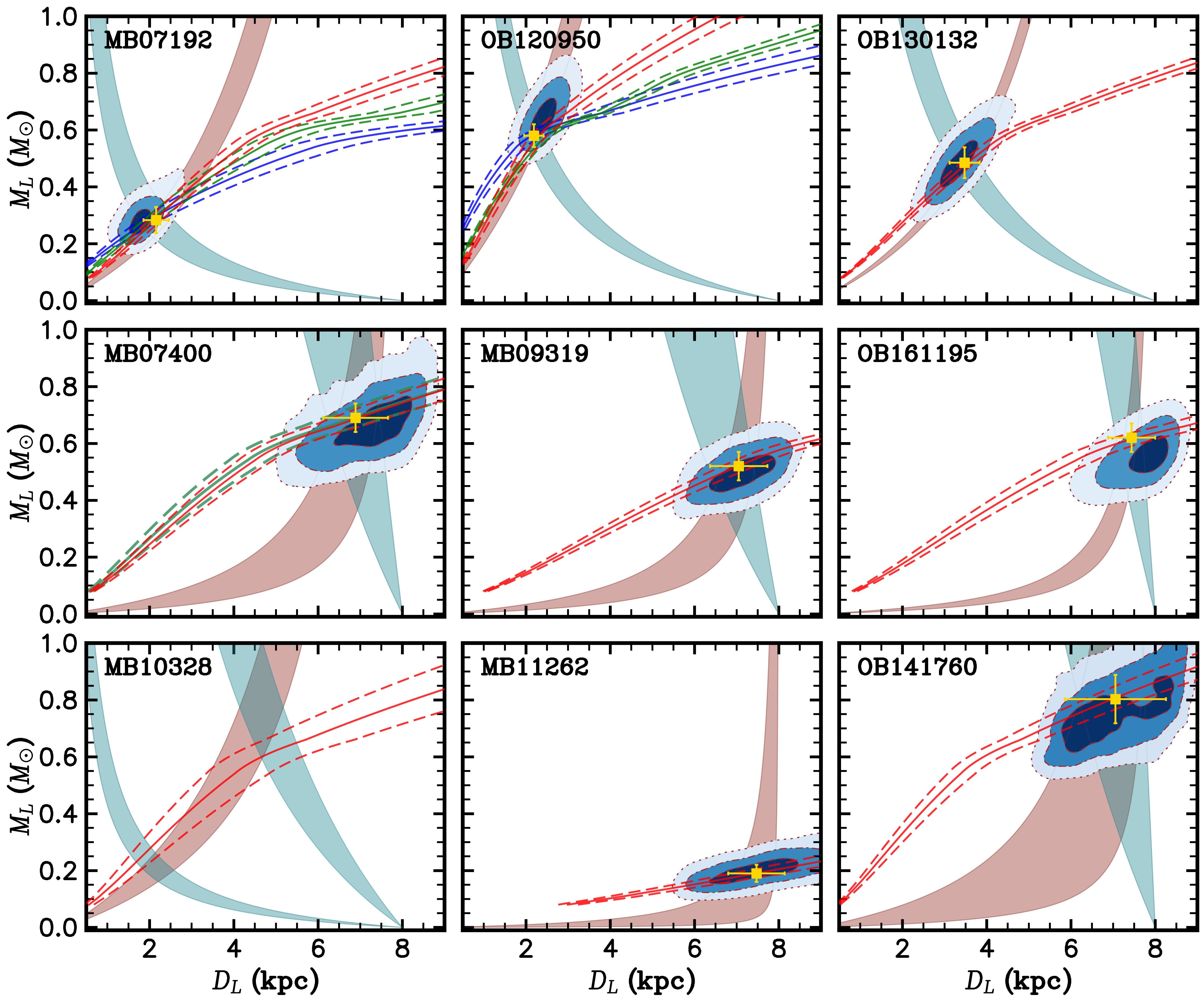}
\centering
\caption{Similar to Figure \ref{fig:MD_grid_simEvents}, but for a representative sample of nine historical events from Table \ref{tab:eesunhong-pylimass}. The top row shows nearby ($< 4$ kpc) events for which a combination of all three mass-distance relations gives an unambiguously measured lens mass and distance. The constraints from $\theta_E$, $\pi_E$, and lens flux are shown as the brown, teal, and red/blue/green curves. The lens flux constraints in these cases come from empirical $V$, $I$, and $K$ band mass-luminosity relations of \cite{delfosse:2000a}. The yellow data with error bar shows the best fit result from image-constrained modeling, and the blue contours give the posterior distribution (central 39\%, 86\%, 99\%) from \texttt{pyLIMASS}. The middle row shows more distant events ($> 4$ kpc) for which the estimate of lens mass and distance is still relatively accurate but less precise. The bottom row shows problematic events for which there are degeneracies that remain (MB10328; \citealt{vandorou:2025b}), the identification of the lens is dubious (MB11262; \citealt{terry:2025a}), or the significantly overlapping $\theta_E$ and $\pi_E$ relations result in large errors (OB141760; \citealt{rektsini:2025a}). \label{fig:MD_grid}}
\end{figure*}

\indent Figure \ref{fig:pylimass-vs-empirical} shows a comparison between the inferred lens mass and distance from image-constrained light curve modeling and \texttt{pyLIMASS} for the 15 historical events we present in Table \ref{tab:eesunhong-pylimass}. We also include six events from \cite{bachelet:2024a} (in blue) for which they made a similar comparison between light curve modeling and \texttt{pyLIMASS}. The distribution shows good agreement across nearly the full range of masses and distances inferred from both methods.\\
\indent There appears to be a small deviation in $D_L$ at large lens distances, which may indicate a small bias in \texttt{pyLIMASS} toward larger source distances. In fact this was previously pointed out in \cite{bachelet:2024a}, and we confirm the same behavior in our expanded sample. The choice of source distance prior via galactic models likely drives this bias, which is further supported by the trend seen in Figure \ref{fig:individ_events_hist} that we described in Section \ref{subsec:pylimass_results}. As we discussed previously, the expected single-epoch astrometric precision in the GBTDS is sufficient to directly measure the source parallax for many of the bulge sources, if systematics can be controlled to the level of the statistical uncertainties. Direct estimates of the distances to the sources should mitigate the dependence on the prior on the source distance from a galactic model, and should also minimize any bias in the lens distances derived from \texttt{pyLIMASS}. \\
\indent We note three important aspects of our \texttt{pyLIMASS} vs. light curve modeling comparisons. First, there is one event in common between our sample and the \cite{bachelet:2024a} sample, event OGLE-2019-BLG-0960. This event, originally published by \cite{yee:2021a}, has subsequent high-resolution followup imaging by \cite{zhang:2025a} which further constrains the lens system physical properties. While the \cite{bachelet:2024a} \texttt{pyLIMASS} run uses observables from \cite{yee:2021a}, the \texttt{pyLIMASS} run in our study uses observables presented in the \cite{zhang:2025a} follow-up work. We list all three studies, \cite{yee:2021a}, \cite{bachelet:2024a} (using \texttt{pyLIMASS}), and \cite{zhang:2025a} (using high-res followup imaging), along with our updated \texttt{pyLIMASS} results in Table \ref{tab:eesunhong-pylimass}. Secondly, three out of the six events presented in \cite{bachelet:2024a} do not have high-resolution followup imaging constraints applied to the light curve modeling, however we choose to keep these data points in Figure \ref{fig:pylimass-vs-empirical} for completeness. Lastly, for MOA-2010-BLG-328 \citep{furusawa:2013a, vandorou:2025b}, two solutions for the lens system physical parameters exists even after the high-resolution followup imaging. Therefore we include both solutions presented by \cite{vandorou:2025b} in Table \ref{tab:eesunhong-pylimass} and Figure \ref{fig:pylimass-vs-empirical}.\\
\indent In Table \ref{tab:eesunhong-pylimass} and Figure \ref{fig:pylimass-vs-empirical} we have shown over a dozen examples of follow-up analyses that leveraged a combination of mass-distance relations to get relatively precise estimates of lens system physical properties. Figure \ref{fig:MD_grid} shows a grid of mass-distance plots for a representative sample of nine events included in Table \ref{tab:eesunhong-pylimass}. The robustness of the final constraints on $M_L$ and $D_L$ for these events varies significantly, and are illustrative of the range of possible outcomes we can expect with Roman.\\
\indent If at least two of the three mass-distance relations ($\theta_E, \pi_E$, and/or lens flux) can be measured with reasonable precision, then the mass and distance of the lens can be unambiguously determined. This is the case for the events in the top and middle rows of Figure \ref{fig:MD_grid} (MB07192, OB120950, OB130132, MB07400, MB09319, OB161195).\\
\indent On the other hand, the bottom row of Figure \ref{fig:MD_grid} shows events for which the final interpration is more problematic. For MB10328 (lower-left panel of Figure \ref{fig:MD_grid}), despite a $K-$band lens detection with Keck AO, there remains two degenerate solutions (e.g. large $\pi_E$ and small $\pi_E$) between a nearby late M dwarf and a more distant early M dwarf \citep{vandorou:2025b}. The prospect of breaking this degeneracy involves obtaining additional observations of the lens in shorter passbands. The problematic interpretation of MB11262 (lower-middle panel of Figure \ref{fig:MD_grid}) is not due to any degeneracy, but it relies on the detected candidate to be the true lens host. Although \cite{terry:2025a} assume the detected candidate is the true lens host, they caution that the probability of the candidate being the true lens is 5.5$\times$ less likely than the candidate being an unrelated field star. This is largely due to the very-high velocity that is inferred if the candidate is the true lens (nearly a hyper-velocity system). Lastly, OB141760 (lower-right panel of Figure \ref{fig:MD_grid}) shows an example of an event with poorly-constrained $M_L$ and $D_L$. In this case, this is not due to any inherent degeneracy or ambiguities in the interpretation, but rather the fact that the uncertainties on $\theta_{\rm E}$ and $\pi_{\rm E}$ are both relatively large, and thus so is the allowed region where they overlap. In this case, the fact that the mass-distance constraint from $K$-band lens flux measurement is relatively shallow where it intersects this allowed region results in a larger fractional uncertainty in $D_L$ than $M_L$ ($\sim 20\%$ versus $\sim 10\%$).

%----------------Discussion and Conclusions------------------------------------------------------------------------------------------------------------

\section{Discussion and Conclusion} \label{sec:conclusion}
In under two years from now, the \textit{Nancy Grace Roman Space Telescope} will be conducting the first ever dedicated space-based microlensing survey to search for cold wide-orbit exoplanets, free-floating planets, and isolated compact objects toward the galactic bulge. As part of mission planning, NASA has established several major science requirements for the microlensing survey. We have investigated one such science requirement that has not been verified until now; \textit{The Roman Space Telescope shall be capable of determining the masses of, and distances to, host stars of 40\% of the detected planets with a precision of 20\% or better}.\\
\indent We have shown, through analysis of simulated \textit{Roman} microlensing events, that the survey can achieve lens mass and distance measurement uncertainties of less than 20\% for nearly half of the detected stellar lens events, and over 40\% for all detected events including VLM and WD hosts. The primary assumptions underlying these results are as follows:

\begin{enumerate}
    \item We adopt the GBTDS Overguide observing strategy as recommended by the \textit{ROTAC} report \citep{rotac:2025a}.
    \item We adopt the fiducial modified \cite{cassan:2012a} mass function to create the sample of simulated planetary microlensing events detected by \textit{Roman} during the RGBTDS.
    \item The \texttt{SynthPop} Galactic model has been used to generate prior probability distributions of source distances for each simulated event.
\end{enumerate}

\indent We have verified that the GBTDS is expected to meet the lens mass and distance measurement requirement, however there are additional factors that may contribute to an overall increase in the final fractions. For instance, utilizing an astrometric microlensing measurement for these events, particularly those with more massive lens hosts, can yield higher precision estimates of the lens mass and distance. Further, as we described in detail in Section \ref{subsec:pylimass_results}, \textit{Roman} may have the sensitivity to directly measure trigonometric parallaxes for source stars in the bulge. This can deliver direct estimates of source distances and mitigate the need to use galactic model priors. Both of these prospects will work to increase the total fraction of events with lens masses and distances measured to 20\% precision or better. Therefore we consider our results presented in Table \ref{tab:pylimass_requirement} as conservative. \\
\indent To study the new algorithm \texttt{pyLIMASS}, we compared the derived lens mass and distance for 15 historical microlensing events that have high-resolution followup imaging from \textit{Keck} and \textit{HST}. We find good agreement (largely within ${\sim}1\sigma$) between the \texttt{pyLIMASS}-estimated mass and distance compared to image-constrained light curve modeling results. This means we can expect \textit{Roman} mass and distance measurements to be robust against the choice of either (or both) methodologies.

\startlongtable
\begin{deluxetable*}{@{\extracolsep{2pt}}lcccccccccc}
\tablecaption{Estimates of lens mass $M_L$, lens distance $D_L$, and source distance $D_S$ for historical events \label{tab:eesunhong-pylimass}}
\setlength{\tabcolsep}{3.0pt}
\tablewidth{\columnwidth}
\tablehead
{
\colhead{}&
  \multicolumn{5}{c}{\hspace{2cm}Light curve only}&
  \multicolumn{3}{c}{\hspace{-0.5cm}Light curve+Imaging} \\
\cline{3-5} \cline{6-8} \cline{9-11}
\colhead{\hspace{-2.5cm}Event} & \colhead{Ref.} & \colhead{$M_L$ [$M_{\odot}$]} & \colhead{$D_L$ [kpc]} & \colhead{$D_S$ [kpc]} & \colhead{$M_L$ [$M_{\odot}$]} & \colhead{$D_L$ [kpc]} & \colhead{$D_S$ [kpc]}
}
\startdata
OGLE-2003-BLG-235 & \cite{bond:2004a} & $0.36^{+0.03}_{-0.28}$ & $5.20^{+0.20}_{-2.90}$ & $-$ & $-$ & $-$ & $-$ \\
{} & \cite{bennett:2006a} & $-$ & $-$ & $-$ & $0.63^{+0.07}_{-0.09}$ & $5.80^{+0.60}_{-0.70}$ & $-$\\
{} & \cite{bhattacharya:2023a} & $0.64 \pm 0.06$ & $5.61 \pm 0.64$ & $8.57 \pm 1.42$ & $0.56 \pm 0.06$ & $5.26 \pm 0.71$ & $8.57 \pm 1.42$\\
{} & \texttt{pyLIMASS} & $4.37^{+5.19}_{-2.58}$ & $7.67^{+0.53}_{-0.72}$ & $8.41^{+0.80}_{-0.85}$ & $0.53^{+0.08}_{-0.07}$ & $5.36^{+0.62}_{-0.66}$ & $8.75^{+1.14}_{-1.17}$ \\
\hline
OGLE-2005-BLG-071 & \cite{dong:2009a} & $-$ & $-$ & $-$ & $0.46 \pm 0.04$ & $3.2 \pm 0.4$ & $8.6$ \\
{} & \cite{bennett:2020a} & $0.47^{+0.36}_{-0.22}$ & $0.72^{+0.81}_{-0.48}$ & $8.76 \pm 1.44$ & $0.43 \pm 0.04$ & $3.46 \pm 0.33$ & $9.28 \pm 1.17$\\
{} & \texttt{pyLIMASS} & $0.43^{+2.77}_{-0.11}$ & $3.16^{+4.10}_{-0.62}$ & $9.36^{+0.63}_{-0.62}$ & $0.36^{+0.06}_{-0.04}$ & $2.82^{+0.47}_{-0.34}$ & $9.29^{+1.27}_{-1.29}$ \\
\hline
MOA-2007-BLG-192 & \cite{kubas:2012a} & $0.08^{+0.02}_{-0.01}$ & $0.70^{+0.21}_{-0.12}$ & $-$ & $0.08^{+0.02}_{-0.01}$ & $0.66^{+0.10}_{-0.07}$ & $-$\\
{} & \cite{terry:2024a} & $0.34^{+0.47}_{-0.19}$ & $1.19^{+1.20}_{-0.67}$ & $7.05 \pm 1.17$ & $0.28 \pm 0.04$ & $2.16 \pm 0.30$ & $7.05 \pm 1.16$\\
{} & \texttt{pyLIMASS} & $0.12^{+0.06}_{-0.03}$ & $0.68^{+0.27}_{-0.16}$ & $5.16^{+0.76}_{-0.97}$ & $0.28^{+0.05}_{-0.04}$ & $1.79^{+0.23}_{-0.16}$ & $6.08^{+0.83}_{-0.70}$ \\
\hline
MOA-2007-BLG-400 & \cite{dong:2009b} & $<0.7$ & $4-8$ & $-$ & $-$ & $-$ & $-$\\
{} & \cite{bhattacharya:2021a} & $0.38 \pm 0.22$ & $6.25 \pm 0.55$ & $7.76^{+0.98}_{-0.87}$ & $0.69 \pm 0.04$ & $6.89 \pm 0.77$ & $7.76^{+0.98}_{-0.87}$\\
{} & \texttt{pyLIMASS} & $0.53^{+0.61}_{-0.29}$ & $6.15^{+1.00}_{-1.27}$ & $7.89^{+0.70}_{-0.84}$ & $0.66^{+0.09}_{-0.08}$ & $7.40^{+0.60}_{-1.00}$ & $8.64^{+0.78}_{-1.25}$ \\
\hline
MOA-2008-BLG-379 & \cite{suzuki:2014a} & $0.56^{+0.24}_{-0.27}$ & $3.3^{+1.3}_{-1.2}$ & $8.0$ & $-$ & $-$ & $-$\\
{} & \cite{bennett:2024a} & $0.70 \pm 0.31$ & $4.0 \pm 1.5$ & $7.8 \pm 1.3$ & $0.43 \pm 0.07$ & $3.44 \pm 0.53$ & $7.8 \pm 1.3$\\
{} & \texttt{pyLIMASS} & $6.99^{+8.56}_{-4.86}$ & $6.82^{+0.91}_{-1.23}$ & $7.84^{+0.72}_{-0.80}$ & $0.56 \pm 0.08$ & $4.55^{+0.70}_{-0.65}$ & $8.35^{+0.97}_{-1.03}$\\
\hline
MOA-2009-BLG-319 & \cite{miyake:2011a} & $0.38^{+0.34}_{-0.18}$ & $6.10^{+1.10}_{-1.20}$ & ${\sim}8.0$ & $-$ & $-$ & $-$\\
{} & \cite{terry:2021a} & $0.33^{+0.40}_{-0.18}$ & $6.75 \pm 0.85$ & $8.25 \pm 0.86$ & $0.52 \pm 0.05$ & $7.05 \pm 0.71$ & $8.25 \pm 0.86$\\
{} & \texttt{pyLIMASS} & $1.58^{+0.73}_{-0.55}$ & $7.59^{+0.59}_{-0.47}$ & $8.22^{+0.60}_{-0.57}$ & $0.47^{+0.07}_{-0.08}$ & $6.83^{+0.76}_{-0.75}$ & $8.26^{+0.99}_{-1.01}$ \\
\hline
MOA-2010-BLG-328 & \cite{furusawa:2013a} & $0.11 \pm 0.01$ & $0.81 \pm 0.10$ & $8.0 \pm 0.3$ & $-$ & $-$ & $-$\\
{} & \cite{vandorou:2025b} & $-$ & $-$ & $-$ & $0.24 \pm 0.02$ & $1.81 \pm 0.21$ & $7.79 \pm 0.52$\\
{} & \cite{vandorou:2025b} & $-$ & $-$ & $-$ & $0.66 \pm 0.05$ & $4.41 \pm 0.26$ & $7.68 \pm 0.58$\\
{} & \texttt{pyLIMASS} & $9.84^{+6.70}_{-4.80}$ & $6.90^{+1.43}_{-1.95}$ & $7.43^{+1.45}_{-2.20}$ & $0.62^{+0.11}_{-0.10}$ & $5.03^{+0.76}_{-0.82}$ & $10.78^{+1.68}_{-2.47}$ \\
\hline
MOA-2010-BLG-477 & \cite{bachelet:2012a} & $0.67^{+0.33}_{-0.13}$ & $2.30 \pm 0.60$ & $8.0 \pm 1.2$ & $-$ & $-$ & $-$\\
{} & \cite{blackman:2021a} & $-$ & $-$ & $-$ & $0.53 \pm 0.11$ & $1.99 \pm 0.35$ & $7.8 \pm 1.3$\\
{} & \texttt{pyLIMASS} & $6.66^{+8.15}_{-3.99}$ & $6.09^{+1.12}_{-1.49}$ & $7.90^{+0.74}_{-0.70}$ & $0.61 \pm 0.10$ & $2.16 \pm 0.30$ & $9.01^{+0.06}_{-0.07}$ \\
\hline
MOA-2011-BLG-262 & \cite{bennett:2014a} & $-$ & $-$ & $-$ & $0.11^{+0.21}_{-0.06}$ & $7.20 \pm 0.80$ & $8.2 \pm 1.0$\\
{} & \cite{terry:2025a} & $0.18^{+0.15}_{-0.07}$ & $7.25 \pm 0.75$ & $7.93 \pm 0.98$ & $0.19 \pm 0.03$ & $7.49 \pm 0.91$ & $7.93 \pm 0.98$\\
{} & \texttt{pyLIMASS} & $1.20^{+1.19}_{-0.63}$ & $7.75^{+0.67}_{-0.59}$ & $7.89^{+0.65}_{-0.64}$ & $0.18 \pm 0.04$ & $7.42^{+1.04}_{-0.75}$ & $7.90^{+1.09}_{-0.80}$ \\
\hline
OGLE-2012-BLG-0563 & \cite{fukui:2015a} & $-$ & $-$ & $-$ & $0.34^{+0.12}_{-0.20}$ & $1.30^{+0.60}_{-0.80}$ & $9.10^{+0.90}_{-1.10}$\\
{} & \cite{bennett:2024b} & $0.18^{+0.16}_{-0.08}$ & $4.19 \pm 0.56$ & $7.45 \pm 1.28$ & $0.80 \pm 0.03$ & $5.49^{+0.58}_{-0.56}$ & $8.48 \pm 1.14$\\
{} & \texttt{pyLIMASS} & $0.36^{+0.10}_{-0.13}$ & $1.39^{+0.38}_{-0.46}$ & $9.06^{+0.67}_{-0.69}$ & $0.78^{+0.13}_{-0.11}$ & $5.57^{+0.51}_{-0.44}$ & $8.40^{+0.80}_{-0.71}$ \\
\hline
OGLE-2012-BLG-0950 & \cite{koshimoto:2017a} & $0.56^{+0.12}_{-0.16}$ & $3.0^{+0.8}_{-1.1}$ & $8.0 \pm 1.6$ & $-$ & $-$ & $-$ \\
{} & \cite{bhattacharya:2018a} & $0.48^{+0.88}_{-0.24}$ & $3.06^{+1.40}_{-1.60}$ & $8.2 \pm 1.5$ & $0.58 \pm 0.04$ & $2.19 \pm 0.23$ & $8.2 \pm 1.5$\\
{} & \texttt{pyLIMASS} & $0.61^{+0.22}_{-0.15}$ & $3.29^{+0.71}_{-0.54}$ & $8.22^{+1.49}_{-1.59}$ & $0.65^{+0.07}_{-0.06}$ & $2.42^{+0.25}_{-0.23}$ & $8.37^{+0.82}_{-0.95}$\\
\hline
OGLE-2013-BLG-0132 & \cite{mroz:2017a} & $0.54^{+0.30}_{-0.23}$ & $3.90^{+1.5}_{-1.3}$ & $-$ & $-$ & $-$ & $-$\\
{} & \cite{rektsini:2024a} & $0.59^{+0.40}_{-0.38}$ & $3.75 \pm 1.24$ & $-$ & $0.50 \pm 0.05$ & $3.48 \pm 0.36$ & $7.41 \pm 0.71$\\
{} & \texttt{pyLIMASS} & $0.40^{+0.31}_{-0.22}$ & $4.59^{+1.05}_{-1.45}$ & $7.70^{+0.49}_{-0.59}$ & $0.43 \pm 0.09$ & $3.10^{+0.45}_{-0.44}$ & $6.85^{+0.84}_{-0.79}$ \\
\hline
OGLE-2014-BLG-1760 & \cite{bhattacharya:2016a} & $0.54^{+0.30}_{-0.23}$ & $3.90^{+1.5}_{-1.3}$ & $8.69^{+1.45}_{-0.76}$ & $-$ & $-$ & $-$\\
{} & \cite{rektsini:2025a} & $-$ & $-$ & $-$ & $0.80 \pm 0.10$ & $7.06 \pm 1.47$ & $10.32^{+1.98}_{-0.72}$\\
{} & \texttt{pyLIMASS} & $0.97^{+0.21}_{-0.20}$ & $11.10^{+1.08}_{-2.46}$ & $13.25^{+1.28}_{-3.16}$ & $0.78^{+0.21}_{-0.16}$ & $7.57^{+2.08}_{-2.52}$ & $11.25^{+2.65}_{-3.16}$ \\
\hline
OGLE-2016-BLG-1195 & \cite{bond17} & $0.37^{+0.38}_{-0.21}$ & $7.20^{+0.80}_{-1.02}$ & $-$ & $-$ & $-$ & $-$\\
{} & \cite{shvartzvald:2017a} & $0.08^{+0.02}_{-0.01}$ & $3.91^{+0.42}_{-0.46}$ & $-$ & $-$ & $-$ & $-$\\
{} & \cite{vandorou:2025a} & $-$ & $-$ & $-$ & $0.62 \pm 0.05$ & $7.45 \pm 0.55$ & $7.8 \pm 1.3$\\
{} & \texttt{pyLIMASS} & $0.60^{+0.08}_{-0.07}$ & $9.23^{+0.53}_{-1.03}$ & $11.20^{+0.72}_{-1.37}$ & $0.60^{+0.08}_{-0.07}$ & $7.97^{+0.68}_{-1.05}$ & $10.81^{+0.86}_{-1.36}$\\
\hline
OGLE-2019-BLG-0960 & \cite{yee:2021a} & $0.50 \pm 0.12$ & $0.98 \pm 0.21$ & $7.56$ & $-$ & $-$ & $-$\\
{} & \cite{bachelet:2024a} & $0.50 \pm 0.10$ & $1.30 \pm 0.30$ & $9.4 \pm 2.7$ & $-$ & $-$ & $-$\\
{} & \cite{zhang:2025a} & $-$ & $-$ & $-$ & $0.40 \pm 0.03$ & $0.92^{+0.09}_{-0.08}$ & $8.0$\\
{} & \texttt{pyLIMASS} & $0.44^{+0.16}_{-0.15}$ & $0.96^{+0.46}_{-0.35}$ & $10.75^{+3.81}_{-3.09}$ & $0.44^{+0.15}_{-0.13}$ & $0.96^{+0.40}_{-0.30}$ & $10.08^{+2.07}_{-2.64}$
\enddata
\end{deluxetable*}

\begin{deluxetable*}{lcc}[!h]
\deluxetablecaption{\texttt{pyLIMASS} Observables \& Priors for the Historical Events\label{tab:observables-priors}}
\tablecolumns{3}
\setlength{\tabcolsep}{13pt}
\tablewidth{\columnwidth}
\tablehead{
\colhead{\hspace{-2.4cm}Event} & \colhead{Observables} & \colhead{Priors}
}
\startdata
OGLE-2003-BLG-235 & $t_E, t_*, V_S, I_S, K_S, A_{V_S}, K_L, \mu_{\textrm{rel,HN}}, \mu_{\textrm{rel,HE}}$ & $D_S = \mathcal{N}(8.6, 1.9)$ kpc\\
\hline
OGLE-2005-BLG-071 & $t_E, t_*, V_S, I_S, K_S, A_{V_S}, K_L, \mu_{\textrm{rel,HN}}, \mu_{\textrm{rel,HE}}, \mu_{\textrm{S,N}}, \mu_{\textrm{S,E}}$ & $D_S = \mathcal{N}(9.3, 2.0)$ kpc\\
\hline
MOA-2007-BLG-192 & $t_E, t_*, V_S, I_S, K_S, A_{V_S}, V_L, I_L, K_L, \theta_*, \mu_{\textrm{rel,HN}}, \mu_{\textrm{rel,HE}}$ & $D_S = \mathcal{N}(7.1, 2.0)$ kpc\\
\hline
MOA-2007-BLG-400 & $t_E, t_*, V_S, I_S, H_S, K_S, A_{V_S}, H_L, K_L, \mu_{\textrm{rel,HN}}, \mu_{\textrm{rel,HE}}$ & $D_S = \mathcal{N}(7.8, 2.0)$ kpc\\
\hline
MOA-2008-BLG-379 & $t_E, t_*, V_S, I_S, K_S, A_{V_S}, V_L, I_L, K_L, \mu_{\textrm{rel,HN}}, \mu_{\textrm{rel,HE}}$ & $D_S = \mathcal{N}(7.8, 2.1)$ kpc\\
\hline
MOA-2009-BLG-319 & $t_E, t_*, V_S, I_S, K_S, A_{V_S}, K_L, \theta_*, \mu_{\textrm{rel,HN}}, \mu_{\textrm{rel,HE}}$ & $D_S = \mathcal{N}(8.3, 1.5)$ kpc\\
\hline
MOA-2010-BLG-328 & $t_E, t_*, V_{S+L}, I_{S+L}, K_S, A_{V_S}, K_L, \mu_{\textrm{rel,HN}}, \mu_{\textrm{rel,HE}}$ & $D_S = \mathcal{N}(7.4, 3.2)$ kpc\\
\hline
MOA-2010-BLG-477$^{*}$ & $t_E, t_*, V_S, I_S, H_S, K_S, A_{V_S}, \mu_{\textrm{rel,G}}, \theta_E$ & $D_S = \mathcal{N}(7.8, 2.1)$ kpc\\
{} & {} & $M_L = \mathcal{N}(0.6, 0.1)\, M_{\odot}$ \\
\hline
MOA-2011-BLG-262 & $t_E, t_*, V_S, I_S, K_S, A_{V_S}, K_L, \theta_*, \mu_{\textrm{rel,HN}}, \mu_{\textrm{rel,HE}}, \mu_{\textrm{S,N}}, \mu_{\textrm{S,E}}$ & $D_S = \mathcal{N}(7.9, 1.8)$ kpc\\
\hline
OGLE-2012-BLG-0563 & $t_E, V_S, I_S, K_S, A_{V_S}, V_L, I_L, K_L, \mu_{\textrm{rel,HN}}, \mu_{\textrm{rel,HE}}$ & $D_S = \mathcal{N}(7.6, 2.4)$ kpc\\
\hline
OGLE-2012-BLG-0950 & $t_E, t_*, V_S, I_S, K_S, A_{V_S}, V_L, I_L, K_L, \theta_*, \mu_{\textrm{rel,HN}}, \mu_{\textrm{rel,HE}}$ & $D_S = \mathcal{N}(8.2, 1.5)$ kpc\\
\hline
OGLE-2013-BLG-0132 & $t_E, t_*, V_S, I_S, K_S, A_{V_S}, K_L, \theta_*, \mu_{\textrm{rel,HN}}, \mu_{\textrm{rel,HE}}$ & $D_S = \mathcal{N}(7.0, 1.4)$ kpc\\
\hline
OGLE-2014-BLG-1760 & $t_E, t_*, V_S, I_S, K_S, A_{V_S}, K_L, \theta_*, \mu_{\textrm{rel,HN}}, \mu_{\textrm{rel,HE}}$ & $D_S = \mathcal{N}(8.3, 2.1)$ kpc\\
\hline
OGLE-2016-BLG-1195 & $t_E, t_*, V_S, I_S, K_S, A_{V_S}, K_L, \theta_*, \mu_{\textrm{rel,HN}}, \mu_{\textrm{rel,HE}}$ & $D_S = \mathcal{N}(7.0, 1.9)$ kpc\\
\hline
OGLE-2019-BLG-0960 & $t_E, V_S, I_S, K_S, A_{V_S}, K_L, \rho_*, \theta_*$ & $D_S = \mathcal{N}(8.0, 2.0)$ kpc\\
\enddata
\tablenotetext{}{\footnotesize{\textbf{$^*$}This event has a WD host, \texttt{pyLIMASS} does not currently support WD isochrones. We constrained the host mass with a prior based on the observed WD mass distribution from \cite{tremblay:2016a}.}}
\end{deluxetable*}

%\begin{acknowledgements}
\section*{Acknowledgments}
\noindent Funding for the Roman Galactic Exoplanet Survey Project Infrastructure Team is provided by the Nancy Grace Roman Space Telescope Project through the National Aeronautics and Space Administration grant 80NSSC24M0022, by The Ohio State University through the Thomas Jefferson Chair for Space Exploration endowment, and by the Vanderbilt Initiative in Data-intensive Astrophysics (VIDA). JPB has been supported by the Australian
Government through the Australian Research Council Discovery Project Grants 240101842, JPB and AC are supported by the SPACE-MLENS ANR grant ANR-24-CE31-3263. MJH Acknowledges support from the Heising-Simons Foundation under grant No. 2022-3542.

%\end{acknowledgements}

\noindent \textit{Software}: Astropy \citep{robitaille:2013a}, eesunhong \citep{bennett:1996a}, genulens \citep{koshimoto:code}, Matplotlib \citep{hunter:2007a}, Numpy \citep{oliphant:2006a}, pyLIMASS \citep{bachelet:2024a}, SynthPop \citep{kluter:2025a}.

\bibliographystyle{aasjournal}
\bibliography{Terry_rges.bib}

\end{document}